\documentclass[a4paper,11pt]{article}
\pdfoutput=1 
\usepackage{amssymb,mathrsfs}
\usepackage{bbm}
\usepackage{siunitx}
\usepackage{amsmath}
\usepackage{braket}
\usepackage{mathtools}
\usepackage[usenames,dvipsnames,svgnames,table]{xcolor}
\usepackage [utf8] {inputenc}
\usepackage{color}
\usepackage{subcaption}
\usepackage{lmodern}
\usepackage{footnote}
\usepackage{slashed}
 \usepackage{mathrsfs}
 \usepackage[pdftex] {graphicx}
\usepackage{multirow}
\usepackage{jheppub} 
\usepackage{here}
\usepackage{hyperref}
\usepackage[justification=centering, singlelinecheck=false]{caption}
\usepackage[list=true, labelfont=bf, labelformat=brace, position=top]{subcaption}
\newcommand{\eq}{\begin{eqnarray}}
\newcommand{\en}{\end{eqnarray}}

\title{Three-particle Lellouch-L\"uscher formalism in moving frames}

\author[1]{Fabian M\"uller,}
\affiliation[1]{Helmholtz-Institut f\"ur Strahlen- und Kernphysik (Theorie) and Bethe Center for Theoretical Physics, Universit\"at Bonn, 53115 Bonn, Germany}
\emailAdd{f.mueller@hiskp.uni-bonn.de}

\author[2]{Jin-Yi Pang,}
\affiliation[2]{College of Science, University of Shanghai for Science and Technology, Shanghai 200093, China}
\emailAdd{jypang@usst.edu.cn}

\author[1,3]{Akaki Rusetsky,}
\affiliation[5]{Tbilisi State  University,  0186 Tbilisi, Georgia}
\emailAdd{rusetsky@hiskp.uni-bonn.de}

\author[4]{and Jia-Jun Wu}
\affiliation[4]{School of Physical Sciences, University of Chinese Academy of Sciences, Beijing 100049, China}
\emailAdd{wujiajun@ucas.ac.cn}

\abstract{
  A manifestly relativistic-invariant Lellouch-L\"uscher formalism for the
  decays into three identical particles with no two-to-three transitions is proposed. 
  Similarly to~\cite{Muller:2020wjo}, the formalism
  is based on the use of the non-relativistic effective Lagrangians.
  Manifest Lorentz invariance is guaranteed, as in~\cite{Muller:2021uur},
  by choosing the quantization axis
  along the total four-momentum of the three-particle system. A systematic inclusion of
  the higher-order derivative couplings, as well as higher partial waves is addressed.
}

\allowdisplaybreaks
\begin{document}
\maketitle
\flushbottom

\section{Introduction}

The study of the three-particle systems on the lattice has attracted much attention
in the recent decade~\cite{Kreuzer:2008bi,Kreuzer:2009jp,Kreuzer:2010ti,Kreuzer:2012sr,Briceno:2012rv,Polejaeva:2012ut,Jansen:2015lha,Hansen:2014eka,Hansen:2015zta,Hansen:2015zga,Hansen:2016fzj,Guo:2016fgl,Sharpe:2017jej,Guo:2017crd,Guo:2017ism,Meng:2017jgx,Briceno:2017tce,Hammer:2017uqm,Hammer:2017kms,Mai:2017bge,Guo:2018ibd,Guo:2018xbv,Klos:2018sen,Doring:2018xxx,Briceno:2018mlh,Briceno:2018aml,Mai:2019fba,Guo:2019ogp,Guo:2020spn,Blanton:2019igq,Pang:2019dfe,Jackura:2019bmu,Briceno:2019muc,Romero-Lopez:2019qrt,Konig:2020lzo,Brett:2021wyd,Hansen:2020zhy,Blanton:2020gha,Blanton:2020jnm,Pang:2020pkl,Hansen:2020otl,Romero-Lopez:2020rdq,Blanton:2020gmf,Muller:2020vtt,Blanton:2021mih,Muller:2021uur,Beane:2007es,Detmold:2008fn,Detmold:2008yn,Blanton:2019vdk,Horz:2019rrn,Culver:2019vvu,Fischer:2020jzp,Alexandru:2020xqf,Romero-Lopez:2018rcb,Blanton:2021llb,Mai:2021nul,Mai:2018djl,Blanton:2021eyf,Sadasivan:2021emk,Garofalo:2022pux}.
These studies imply, first and foremost, the measurement of the three-body spectrum
which is further analyzed by using the quantization condition
(an equation that connects the
finite-volume energy spectrum with the infinite-volume observables in
the three-particle system). The parameters, characterizing the three-body interactions
in the infinite volume, could be extracted in a result of this analysis.
In the literature, one finds three conceptually equivalent formulations
of the three-body quantization condition: the so-called
RFT~\cite{Hansen:2014eka, Hansen:2015zga},
NREFT~\cite{Hammer:2017uqm, Hammer:2017kms} and
FVU~\cite{Mai:2017bge,Mai:2018djl} approaches. Note that a Lorentz-invariant
formulation of the NREFT approach was suggested recently~\cite{Muller:2021uur}.
We shall be using this approach in what follows. For more information
on the subject, we refer the reader to the two recent reviews on the
subject~\cite{Hansen:2019nir,Mai:2021lwb}.

Furthermore, a three-body analog of the
Lellouch-L\"uscher (LL) formula, which relates the three-body decay amplitudes,
measured in a finite and in the infinite volume, has been
derived~\cite{Muller:2020wjo,Hansen:2021ofl} in the NREFT and RFT settings, respectively.
The Ref.~\cite{Muller:2020wjo} was more a proof of principle where the relation between
the three-body decay amplitudes in a finite and in the infinite volume has been worked out
only at the leading order in the EFT expansion. The technical details have been left for future work. The aim of the present paper is to complete the derivation, given in
Ref.~\cite{Muller:2020wjo}, in order to obtain a general three-body
LL formula, and to carry out the comparison with the
findings of Ref.~\cite{Hansen:2021ofl}. A number of non-trivial issues have to be addressed
to fill this gap. Namely,

\begin{itemize}

\item[i)] Only S-wave contributions were allowed in the formula given in Ref.~\cite{Muller:2020wjo}. In some cases, the contributions from higher partial waves may be essential.\footnote{An obvious example here is the decay $\omega\to 3\pi$. The decay width of this resonance is rather small, and the pole lies very close to the real axis. Hence, the formalism considered in this paper can be directly applied to this case.}
  A systematic inclusion of the higher partial waves has to be performed.

\item[ii)] The role of the derivative couplings could be substantial, especially if one
  considers decays into light particles (for example, pions). As already mentioned
  in~\cite{Muller:2020wjo}, if the derivative vertices describing the three-particle decays
  are included, one does not end up with a single
  LL factor (in difference to the two-body case). Albeit a relatively
  straightforward task, an explicit
  form of the  LL relation in this case should be still worked out.

\item[iii)] Even if the formula given in Ref.~\cite{Muller:2020wjo} has the relativistic appearance (for example, the single-particle energies are given by the relativistic expression
  $w({\bf p})=\sqrt{m^2+{\bf p}^2}$), it is valid only in the center-of-mass (CM) frame and at the leading order in the EFT expansion. Including higher-dimensional
  operators in the Lagrangian leads to the proliferation of the number of the effective couplings that have to be determined from the fit to the lattice data. To this end, it would be useful to use data from the sectors with different total momenta. This is however possible, if and only if the approach is manifestly Lorentz-invariant.\footnote{Here, we would like to mention that the manifest relativistic invariance of the RFT approach comes at the cost of
    imposing a cutoff on the spectator momentum in the three-body equation,
    which is of order of the particle mass. Increasing the cutoff beyond some critical value is not allowed as it would lead to the spurious singularities in the amplitude. For a more detailed discussion of the issue, we refer the reader to~\cite{Muller:2021uur}.}

\end{itemize}

In order to achieve the goal stated above, we merge the NREFT derivation of
the LL framework, given in Ref.~\cite{Muller:2020wjo}, with the
manifestly invariant three-particle setting of Ref.~\cite{Muller:2021uur}. The layout
of the paper is as follows. In Sect.~\ref{sec:basic} we list the main relations that
define a Lorentz-invariant three-particle quantization condition. Here, we also
display the effective Lagrangian that describes the three-particle decays at tree level
in a relativistic-invariant fashion. We remind the reader that the decays are assumed
to proceed via a different mechanism (e.g., through the weak or electromagnetic
interactions) than the formation of the colorless bound states from quarks and gluons.
For this reason, the masses of all particles (hadrons) are real, and
the effective Lagrangian can be written down in terms of the fields of all hadrons,
participating the reaction.\footnote{Note that this differs from the case of the QCD resonances, like $a_1(1260)$ or the Roper resonance, which correspond to a pole in the complex energy plane.} Furthermore, in Sect.~\ref{sec:derivation} we give a detailed derivation of the
LL framework in a manifestly Lorentz-invariant setting.
Finally, Sect.~\ref{sec:concl} contains our main result and conclusions.

\section{Relativistic invariant framework in the three-particle sector}

\label{sec:basic}

\subsection{The Lagrangian}

In the non-relativistic effective theory time and space coordinates are treated differently. For this reason, this theory is not manifestly Lorentz-invariant.
 In Ref. ~\cite{Muller:2021uur}
the invariance was achieved by using the following trick. At the first stage, the quantization axis was chosen along the arbitrary unit four-vector $v^\mu$.
Using this vector, all expressions can be rewritten in a manifestly invariant form.
This however does not suffice, since the presence of an ``external'' vector $v^\mu$ signals
the breakdown of the Lorentz invariance. Only at the next stage, when $v^\mu$ is fixed
in terms of the external momenta in a given process, the relativistic invariance is restored.\footnote{With the choice of an arbitrary quantization axis $v^\mu$,
  one has to define, what does
  one now mean under time and space coordinates. Let $\underline{\Lambda}$ be the Lorentz
  transformation $\underline{\Lambda} v=v_0$, where $v^\mu_0=(1,{\bf 0})$. Further, let $\underline{\Lambda} x=x'$.
  Then, ${x'}^0$ and ${\bf x}'$ are set to play the role of time and space coordinates.}

Below, we shall briefly sketch the formalism of Ref. ~\cite{Muller:2021uur}.
It is convenient to work in
the particle-dimer picture, which
has proven to be very useful for the derivation
of the Faddeev equation in the infinite as well as in a finite volume. A dimer corresponds
to an auxiliary field (an integration variable) introduced in this Lagrangian and, thus, not
necessarily to a physical bound state of two particles, which may or may not exist in a
channel with given quantum numbers. The Lagrangian that describes the three-particle system in question is written down in a following compact form (more details and derivation can be found in Ref.~\cite{Muller:2021uur}):
\eq\label{eq:L}
\mathscr{L}&=&\phi^\dagger 2w_v(i(v\partial)-w_v)\phi
+\sum_{\ell m}\sigma_\ell T_{\ell m}^\dagger T_{\ell m}
+\sum_{\ell m} (T_{\ell m}^\dagger O_{\ell m}+\mbox{h.c.})
\nonumber\\[2mm]
&+&4\pi\sum_{\ell m}\sum_{\ell'm'}\sum_{LL'}\sum_{JM} T_{\ell' m'}^\dagger
\left(\mathscr{Y}^{L'\ell'}_{JM}(\underline{\bf w},m')\phi^\dagger\right)
T_{JL'L}^{\ell'\ell}(\Delta,\mbox{\raisebox{.05cm}{$\stackrel{\leftarrow}{\Delta}$}}_T,\mbox{\raisebox{.05cm}{$\stackrel{\rightarrow}{\Delta}$}}_T)
\left((\mathscr{Y}^{L\ell}_{JM}(\underline{\bf w},m))^*\phi\right)T_{\ell m}\, .
\nonumber\\
\en
The notations in the above (rather compact) formula should be explained in detail. This is done in what follows.
First, $\phi$ denotes a non-relativistic
field operator for the scalar field with the mass $m$,
and $\partial^\mu=(\partial_0,\boldsymbol{\nabla})$
(we remind the reader that in this
paper we consider a system that consists of three identical spinless particles
only and assume that the transitions between the sectors with a different
number of particles are forbidden).
The quantity $w_v=\sqrt{m^2+\partial^2-(v\partial)^2}$
corresponds to the on-shell energy of this particle in the quantization scheme defined
by the vector $v^\mu$ and reduces to a familiar expression $w=\sqrt{m^2-\boldsymbol{\nabla}^2}$
in the rest frame. For simplicity, we shall assume from the beginning that the unit
vector $v^\mu$ is directed along the total four-vector $K^\mu$ of the three-particle system.

Furthermore, $T_{\ell m}$ is a dimer field with a spin $\ell$ and projection $m=-\ell,\cdots,\ell$. This field is constructed as follows. One starts from the tensor fields
with $\ell$ indices $T_{\mu_1\cdots\mu_\ell}$. These fields are symmetric under a permutation of each two indices, traceless in each pair of indices and obey the constraint
\eq\label{eq:A3}
v^{\mu_i}T_{\mu_1\cdots\mu_\ell}=0\, ,\quad\quad i=1,\cdots,\ell\, .
\en
Next, let $\underline{\Lambda}$ be a matrix of Lorentz transformation that transforms $v^\mu$ into
$v_0^\mu$:
\eq
\underline{\Lambda}(v)^{00}=v^0\, ,\quad\quad 
\underline{\Lambda}(v)^{0i}=-\underline{\Lambda}(v)^{i0}=v^i\, ,\quad\quad
\underline{\Lambda}(v)^{ij}=-\delta^{ij}-\frac{v^iv^j}{v^0+1}\, ,
\en
or,
\eq\label{eq:Lambda_u}
\underline{\Lambda}(v)^{\mu\nu}=g^{\mu\nu}
-\frac{v^\mu v^\nu}{1+(vv_0)}
-\frac{v_0^\mu v_0^\nu}{1+(vv_0)}
+\frac{v^\mu v_0^\nu+v_0^\mu v^\nu}{1+(vv_0)}\,(vv_0)
-(v^\mu v_0^\nu-v_0^\mu v^\nu)\, .
\en
Then,
\eq\label{eq:lmmu}
T_{\ell m}=\sum_{\mu_1\cdots\mu_\ell}(c^{-1})^{\ell m}_{\mu_1\cdots\mu_\ell}
\underline{\Lambda}_{\nu_1}^{\mu_1}\cdots\underline{\Lambda}_{\nu_\ell}^{\mu_\ell}
T^{\nu_1\cdots\nu_\ell}\, ,
\en
see Eqs. (3.9) and (3.10) of Ref.~\cite{Muller:2021uur}.\footnote{Note that, despite
  manifestly covariant notations used in Eq.~(\ref{eq:Lambda_u}), the quantity
  $\underline{\Lambda}$ is {\em not} a second-rank Lorentz tensor, since under the Lorentz transformations, the vector $v_0$ stays put.} The matrix elements of
$c$, which can be trivially derived,
are purely of group-theoretical origin
(see Appendix~\ref{app:c}).
A general expression for an arbitrary $\ell$ is rather
clumsy and will not be displayed in the main text.
Note also that in Ref.~\cite{Muller:2021uur} the Lagrangian was written down
in terms of the tensor fields $T_{\mu_1\cdots\mu_\ell}$. 
Owing to the orthogonality of the matrices $c$ and $\Lambda$,
the sum over the indices $\mu_1,\cdots,\mu_\ell$ in the Lagrangian
can be readily rewritten as a sum over the indices $\ell,m$.

The second term in the Lagrangian describes the ``free dimer.'' The value of the
parameter $\sigma_\ell=\pm 1$ is fixed by the sign of the two-body scattering length.
The third term describes the interaction of a dimer with a spin $\ell$ coupled to a pair of particles
in the state with an angular momentum $\ell$ (the sum over all $\ell$ is carried out at the
end). The two-particle operators $O_{\ell m}$ take the form, similar to Eq.~(\ref{eq:lmmu}):
\eq\label{eq:lmmu-O}
O_{\ell m}=\sum_{\mu_1\cdots\mu_\ell}(c^{-1})^{\ell m}_{\mu_1\cdots\mu_\ell}
\underline{\Lambda}_{\nu_1}^{\mu_1}\cdots\underline{\Lambda}_{\nu_\ell}^{\mu_\ell}
O^{\nu_1\cdots\nu_\ell}\, .
\en
Here, the fully symmetric operators $O^{\mu_1\cdots\mu_\ell}$ are traceless in each of two indices
and obey the relation
\eq
v^{\mu_i}O_{\mu_1\cdots\mu_\ell}=0\, ,\quad\quad i=1,\cdots,\ell\, .
\en
These operators are constructed of two fields $\phi$ and the vector $\bar w^\mu_\perp=\bar w^\mu-v^\mu(v\bar w)$,
where $\bar w^\mu=\Lambda^\mu_\nu w^\nu$
and the differential operator $w^\mu$ is given by $w^\mu=v^\mu w_v
+i\partial^\mu_\perp$, where $\partial^\mu_\perp=\partial^\mu-v^\mu(v\partial)$.
Note also that the $\Lambda^\mu_\nu$ differs from $\underline{\Lambda}^\mu_\nu$,
which was introduced above. Namely, $\Lambda^\mu_\nu$ is the Lorentz boost that
renders the total four-momentum of the pair parallel to the vector $v^\mu$.
Thus, once one has chosen $v^\mu$ along the total four-momentum of the system,
in the coordinate space $\Lambda^\mu_\nu$ becomes a differential operator.
For $\ell=0,2,\ldots$ the explicit form of the operator $O^{\mu_1\cdots\mu_\ell}$
(up to an inessential overall normalization) is given by
\eq
O&=&\frac{1}{2}\,\hat f_0\phi^2
\nonumber\\[2mm]
O^{\mu\nu}&=&\frac{3}{2}\,\hat f_2
(\phi(\bar w^\mu_\perp \bar w^\nu_\perp\phi)
-(\bar w^\mu_\perp\phi)(\bar w^\nu_\perp\phi))
\nonumber\\[2mm]
&-&\frac{1}{2}\,\hat f_2(g^{\mu\nu}-v^\mu v^\nu)(\phi(\bar w^\lambda_\perp \bar w_{\perp\lambda}\phi)
-(\bar w^\lambda_\perp\phi)(\bar w_{\perp\lambda}\phi))\, ,
\en
and so on.
Here, $\hat f_\ell$ is a differential operator which can be formally expanded in the Taylor series. For example, for $\ell=0$, 
\eq
\hat f_0\phi^2=f_0^{(0)}\phi^2+\frac{1}{2}\,f_0^{(2)}
(\phi(\bar w^\mu_\perp \bar w_{\perp\mu}\phi)
-(\bar w^\mu_\perp\phi)(\bar w_{\perp\mu}\phi))+\cdots\, .
\en
The coefficients $f^{(0)}_\ell,f^{(2)}_\ell,\ldots$ are related to the
effective-range expansion parameters in the two-body system
(the scattering length, effective range and so on).

The construction for higher values of $\ell$ proceeds straightforwardly
(for identical particles, only even values of $\ell$ are allowed). Namely,
in the free field theory, the matrix element of the operator
$O^{\mu_1\cdots\mu_\ell}$ between the vacuum and the two-particle state is given by
\eq
\langle 0|O^{\mu_1\cdots\mu_\ell}|p_1,p_2\rangle
&=&\frac{\sqrt{2\ell+1} f_\ell(-\bar p_\perp^2)}{N_\ell}\,\tilde O^{\mu_1\cdots\mu_\ell}(\bar p_\perp)\, ,
\nonumber\\[2mm]
\frac{\sqrt{2\ell+1} f_\ell(-\bar p_\perp^2)}{N_\ell}
&=&f_\ell^{(0)}-f_\ell^{(2)}\bar p_\perp^2+\cdots\, ,\quad\quad
N_\ell^{-1}=\frac{2^{\ell/2}\ell!}{A^-_{\ell,\ell-1}\cdots A^-_{\ell,0}}\, .
\en
where the quantities $A^-_{\ell,m}$ are defined in Appendix~\ref{app:c}.
For $\ell=0,2,\ldots$, we have
\eq
\tilde O(\bar p_\perp)&=&1\, ,
\nonumber\\[2mm]
\tilde O^{\mu\nu}(\bar p_\perp)&=&\frac{3}{2}\,
\bar p_\perp^\mu \bar p_\perp^\nu-\frac{1}{2}\,(g^{\mu\nu}-v^\mu v^\nu)\bar p_\perp^2\, ,
\en
and so on. Here, $\bar p_\perp^\mu=\bar p^\mu-v^\mu(v\bar p)$ and $\bar p^\mu=\Lambda^\mu_\nu p^\nu=\frac{1}{2}\,\Lambda^\mu_\nu (p_1-p_2)^\nu$ (we remind the reader that
$p_1^2=p_2^2=m^2$). Note also once more that $\Lambda^\mu_\nu$ depends on the total momentum of the two-particle system $p_1+p_2$.
In order to write down the expression for a generic $O_{\mu_1\cdots\mu_\ell}(\bar p_\perp)$,
one may define the tensors in the three-space (the Latin indices $i_n$ run from 1 to 3):
\eq
P_{i_1\cdots i_\ell}({\bf k})=
N_\ell \sqrt{\frac{4\pi}{2\ell+1}}
c^{\ell m}_{i_1\cdots i_\ell}
\left(\mathscr{Y}_{\ell m}({\bf k})\right)^*\, .
\en
Here,
$\mathscr{Y}_{\ell m}({\bf k})=k^\ell Y_{\ell m}(\theta,\varphi)$, 
$Y_{\ell m}(\theta,\varphi)$ denotes the spherical function, and the coefficients $c$ can be found in Appendix~\ref{app:c}. For $\ell=0,2$ one has
\eq
P({\bf k})=1\, ,\quad\quad
P_{ij}({\bf k})=\frac{3}{2}\,k_i k_j-\frac{1}{2}\,\delta_{ij}{\bf k}^2\, ,\quad\cdots
\en
The general pattern is clear. The operators $\tilde O(\bar p_\perp)$,
$\tilde O^{\mu\nu}(\bar p_\perp)$ are constructed in analogy with $P({\bf k})$,
$P_{ij}({\bf k})$, and so on. Namely, one replaces $k_i$ by $p_\perp^\mu$,
${\bf k}^2$ by $-p_\perp^\mu p_{\perp\mu}$ and $\delta_{ij}$ by
$-g^{\mu\nu}+v^\mu v^\nu$. This prescription is valid for all values of $\ell$. Finally, the operator $O^{\mu_1\cdots\mu_\ell}$ in the coordinate space
can be immediately read off from the momentum-space expression. The
prescription for the off-shell momenta is set by the replacement of
a generic $p^\mu$ by the differential operator $w^\mu$.

The last term of the Lagrangian describes the particle-dimer scattering at tree
level.
For any vector $a^\mu$, we have $\underline{a}^\mu=\underline{\Lambda}^\mu_\nu a^\nu$.
Next,
\eq
\mathscr{Y}_{L\ell}^{JM}({\bf k},m)=\langle L(M-m),\ell m|JM\rangle
\mathscr{Y}_{L(M-m)}({\bf k})\, ,
\en
and $\langle L(M-m),\ell m|JM\rangle$ denotes the pertinent Clebsh-Gordan coefficient. Furthermore,
the quantity $T_{JL'L}^{\ell'\ell}$ is a low-energy polynomial of its arguments, and the
coefficients of the expansion are the low-energy couplings. The operator $\Delta_T$ is a differential operator acting on the dimer field (the arrow shows, on which one it does act).
In the momentum space,
\eq\label{eq:DeltaT}
\Delta_T T_{\ell m}(P)=(P^2-4m^2)T_{\ell m}(P)\, ,
\en
where $P$ denotes the four-momentum of the dimer. It can be expressed as $P=K-p$,
where $K$ is the total momentum of the particle-dimer system and $p$ denotes
the momentum of the spectator, which is assumed to be on shell, i.e., $p^2=m^2$.
Finally, in the momentum space, the operator $\Delta$ can be replaced by $K^2-9m^2$ and, thus, does not depend on the spectator momenta.

\subsection{Matching of the couplings describing particle-dimer scattering}
\label{sec:couplings}

Matching in the dimer framework is a very delicate issue. We have
briefly touched on the issue
in our previous paper~\cite{Muller:2021uur}, and here we would like to extend this discussion. This will hopefully
help to avoid misunderstandings related to the above problem.

To start with, the dimer field in this framework is introduced as a dummy integration
variable in the path integral. At the first glance, a one-to-one mapping of the effective couplings
in the particle-dimer picture and the three-particle picture is guaranteed. Furthermore,
carrying out a perturbative matching of the three-particle $S$-matrix elements imposes
certain constraints on the independent effective couplings, as discussed in Ref.~\cite{Muller:2021uur}:
for example, at the next-to-leading order, as a consequence of the Bose-symmetry,
only one independent coupling
out of three survives in the local vertex that describes particle-dimer
scattering. The number of independent couplings agrees
with the findings of Refs.~\cite{Blanton:2019igq,Blanton:2021eyf}, see
also~\cite{Blanton:2021mih}.

Does this result change, if a shallow physical bound state (a dimer) exists
in some channel? On the one hand, it should not, because introducing a dummy
field does not change the number of relevant parameters. On the other hand,
one has more data to fit now: the $S$-matrix elements both in the three-particle
sector {\em and} in the particle-dimer sector. If these two data sets are independent, there will be less constraints and more independent couplings in the Lagrangian, see again Ref.~\cite{Muller:2021uur}.

As can be seen from the above discussion, the difference boils down to a
question, whether the data from three-particle scattering and particle-dimer
scattering are independent from each other. This is a dynamical question, and
one knows examples of either sort in Nature. For instance, the properties of a
deuteron are very well determined by the low-energy $NN$ scattering amplitude
in the pertinent channel. In other words, the deuteron is a beautiful
example of a {\em hadronic
  molecule} (One arrives at the same result, considering
Weinberg's quantization condition~\cite{Weinberg:1965zz}.). In such a case,
the particle-dimer scattering data are determined by the input from the
three-particle sector and there is no need to consider them separately.

On the other hand,
there are resonances, whose existence can be hardly attributed to the
rescattering. Take the extreme case: if the weak interactions are turned on,
the kaon will be seen as a (very narrow) resonance in the $\pi\pi$ scattering
(for a full analogy with the deuteron, by adjusting quark masses one can even tune the kaon mass to be slightly
below the two-pion threshold).
However, the {\em formation} of the kaon has nothing to do with the rescattering
of pions. The $\pi\pi$ scattering $K$-matrix will contain a pre-existing pole
on the real axis, which will be eventually dressed by the pion loops.
In such a case, the data from the
three-particle sector and the particle-dimer sector are independent, and one needs
more parameters in the Lagrangian to describe the $S$-matrix elements in all
sectors. Expanding the pole term in the two-body $K$ matrix will lead to
the formulae that are {\em formally} the same as in the case of a molecule.
The price to pay for this will be however a very small convergence radius.
This case resembles Chiral Perturbation Theory with/without the $\Delta$-resonance -- integrating out the $\Delta$ leads to the unnaturally large effective
couplings that are almost saturated by the $\Delta$-exchange.

To summarize, the particle-dimer picture provides a very flexible framework
that can be used both in the presence or absence of physical dimers.
Furthermore, when the (shallow)
dimers are present, the case of molecular states should be distinguished
from the one of tightly bound compounds (pre-existing resonances). In the particle-dimer picture, these two cases are merely described by a different number
of the independent
effective couplings, because additional parameters are needed to fix
the position and the residue of the pre-existing pole in the two-body $K$-matrix. One could of
course use the same (overcomplete) set of couplings in all cases, bearing in
mind that if there are no dimers, or the dimers are predominantly molecules,
flat directions emerge in the parameter space, when the fit to the finite-volume
levels is performed.

Last but not least, the above discussion directly applies to the bound
states in the three-particle channel (the trimers). These can also have either
molecular nature, or represent tight compounds defined by a different dynamics.
In the latter case, it could be again advantageous to introduce an elementary
trimer field that will allow one to circumvent the problem with unnaturally
large low-energy couplings.

\subsection{Faddeev equation}

\begin{figure}[t]
  \begin{center}
    \includegraphics[width=9cm]{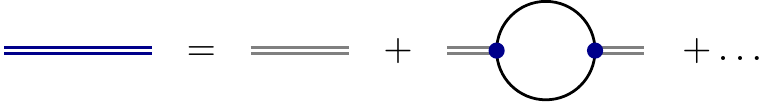}
    \caption{Full dimer propagator, obtained by summing up self-energy insertions to all orders. The double, dashed
      and single lines denote the full dimer propagator, the free dimer propagator given by $-\sigma_\ell^{-1}$,
    and the particle propagator.}
    \label{fig:bubblesummation}
  \end{center}
\end{figure}

Now, we are in a position to write down the Faddeev equation for the particle-dimer scattering and derive the relativistic invariant quantization condition. We start from the infinite-volume case. After summation of the self-energy insertions
(see Fig.~\ref{fig:bubblesummation}), the
dimer propagator can be written down as follows~\cite{Muller:2021uur}:
\eq
i\langle 0|T[T_{\ell m}(x)T^\dagger_{\ell' m'}(y)]|0\rangle=
\delta_{\ell\ell'}\delta_{mm'}\int\frac{d^4P}{(2\pi)^4}\,e^{-iP(x-y)}S_\ell(P^2)\, ,
\en
where
\eq
S_\ell(s)=-\frac{1}{
  \sigma_\ell
  -f^2_\ell(s)\frac{1}{2}\,p^{2\ell}(s)I(s)}
\, ,\quad\quad s=P^2=4(p^2(s)+m^2)\, .
\en
Furthermore,
\eq\label{eq:f2}
16\pi\sqrt{s}\bigl[\sigma_\ell f^{-2}_\ell(s) -\frac{1}{2}\,p^{2\ell}(s)J(s)\bigr]
&=&p^{2\ell+1}(s)\cot\delta_\ell(s)\, ,
\en
and
\eq
I(s)&=&\frac{\sigma(s)}{16\pi^2}\,\ln\frac{\sigma(s)-1}{\sigma(s)+1}\, ,
\quad\quad
\sigma(s)=\biggl(1-\frac{4m^2}{s+i\varepsilon}\biggr)^{1/2}\, ,
\nonumber\\[2mm]
f_\ell(s)&=&f_\ell^{(0)}+f_\ell^{(2)}\biggl(\frac{s}{4}-m^2\biggr)+\cdots\, ,
\en
and $\delta_\ell(s)$ denotes the phase shift in a given partial wave. 
Finally, if $s\geq 4m^2$,
\eq\label{eq:IJ}
I(s)=J(s)+\frac{i\sigma(s)}{16\pi}\, .
\en
The scattering amplitude in a given partial wave $\mathscr{T}_\ell$
is obtained by equipping
the dimer propagator with the endcaps corresponding to the decay of a dimer
into a particle pair. For the process $p_1+p_2\to p_3+p_4$ with the on-shell
particles, the scattering amplitude is written as follows:
\eq
\mathscr{T}(p_1,p_2;p_3,p_4)=
4\pi\sum_{\ell m}\frac{\mathscr{Y}_{\ell m}({\bf\tilde p})\mathscr{Y}_{\ell m}^*({\bf\tilde q})}
{(T^\ell_{\sf tree})^{-1}(s)
  -\frac{1}{2}\,p^{2\ell}(s)I(s)}\, ,
\en
where~\cite{Muller:2021uur}
\eq\label{eq:tilde}
   {\bf\tilde p}=\bar{\bf p}-{\bf v}\frac{\bar{\bf p}{\bf v}-\bar p^0}{{\bf v}^2}\, ,\quad\quad
   {\bf\tilde q}=\bar{\bf q}-{\bf v}\frac{\bar{\bf q}{\bf v}-\bar q^0}{{\bf v}^2}\, ,
   \en
   and
\eq\label{eq:bar} 
\bar p^\mu=\frac{1}{2}\,\Lambda^\mu_\nu(v,u) (p_3-p_4)^\nu\, ,\quad\quad
\bar q^\mu=\frac{1}{2}\,\Lambda^\mu_\nu(v,u) (p_1-p_2)^\nu\, .
\en
Here, $\Lambda(v,u)$ takes the form:
\eq\label{eq:Lambda_uv}
\Lambda(v,u)^{\mu\nu}=g^{\mu\nu}
-\frac{u^\mu u^\nu}{1+(uv)}
-\frac{v^\mu v^\nu}{1+(uv)}
+\frac{u^\mu v^\nu+v^\mu u^\nu}{1+(uv)}\,(uv)
-(u^\mu v^\nu-v^\mu u^\nu)\, .
\en
We remind the reader that $u^\mu=P^\mu/\sqrt{P^2}$ is the unit vector
in the direction of the CM momentum of the pair. Also, the quantity
$\Lambda(v,u)^{\mu\nu}$, in difference to $\underline{\Lambda}(v)^{\mu\nu}$,
{\em is} a second-rank Lorentz-tensor, because both $v^\mu$ and $u^\mu$
transform like Lorentz-vectors. Note also that
$\underline{\Lambda}(v)^{\mu\nu}=\Lambda(v_0,v)^{\mu\nu}$.

The expression (\ref{eq:tilde}) for the vectors ${\bf\tilde p},{\bf\tilde q}$
looks very complicated but, in fact, have a very transparent physical meaning.
In order to show this, note that $(v\bar p)=(v\bar q)=0$. Taking this
into account, it is easy to rewrite Eq.~(\ref{eq:tilde}) in the following form:
\eq\label{eq:tilde1}
         {\bf\tilde p}=\bar{\bf p}+{\bf v}\frac{\bar{\bf p}{\bf v}}{1+v^0}
         -{\bar p^0}{\bf v}\, ,\quad\quad
         {\bf\tilde q}=\bar{\bf q}+{\bf v}\frac{\bar{\bf q}{\bf v}}{1+v^0}
         -{\bar q^0}{\bf v}\, .
   \en
This gives $\tilde p_\mu=\underline{\Lambda}_{\mu\nu}(v)\bar p^\nu$ and 
$\tilde q_\mu=\underline{\Lambda}_{\mu\nu}(v)\bar q^\nu$. Furthermore,
it can be straightforwardly checked that $\tilde p_\mu=\tilde q_\mu=0$.

The full scattering amplitude is obtained by summing $\mathscr{T}_\ell$ over all $\ell$. This summation renders the amplitude relativistic invariant --
the result depends on the Mandelstam variables $s,t$ only.

Next, let $T_{\ell'm',\ell m}(p,q)$
be the particle-dimer scattering amplitude.
The indices $\ell m$/$\ell'm'$ denote the dimer spin and magnetic quantum number
in the initial/final
states, respectively. 
The Faddeev equation takes the matrix form:
\eq\label{eq:Faddeev}
T_{\ell'm',\ell m}(p,q)
&=&Z_{\ell'm',\ell m}(p,q)
\nonumber\\[2mm]
&+&\sum_{\ell''m''}\int^{\Lambda_v}\frac{d^3k_\perp}{(2\pi)^32w_v(k)}\,
Z_{\ell'm',\ell'' m''}(p,k)
S_{\ell''}((K-k)^2)T_{\ell''m'',\ell m}(k,q)\, .\quad\quad\quad
\en
Here, the ``covariant'' cutoff is defined as
\eq
\int^{\Lambda_v}\frac{d^3k_\perp}{(2\pi)^3}\,F(k)
=\int\frac{d^4k}{(2\pi)^3}\,
\delta(k^2-m^2)\theta(\Lambda^2+k^2-(vk)^2)F(k)\, ,
\en
and
\eq\label{eq:driving}
Z_{\ell'm',\ell m}(p,q)
&=&\frac{4\pi\,\left(\mathscr{Y}_{\ell' m'}({\bf\tilde p})\right)^*
  f_{\ell'}(s_p)\,f_\ell(s_q)\mathscr{Y}_{\ell m}({\bf\tilde q})}{2w_v(K-p-q)(w_v(p)+w_v(q)+w_v(K-p-q)-vK-i\varepsilon)}
\nonumber\\[2mm]
&+&4\pi\sum_{LL'}\sum_{JM}\mathscr{Y}_{JM}^{L'\ell'}(\underline{\bf p},m')
T_{JL'L}^{\ell'\ell}(\Delta,\Delta_p,\Delta_q)\left(\mathscr{Y}_{JM}^{L\ell}(\underline{\bf q},m)\right)^*\, .
\en
In the above expression,
${\bf\tilde p},{\bf\tilde q}$ are defined as follows:
\eq
\tilde p^\mu&=&\frac{1}{2}\,\underline{\Lambda}^{\mu\nu}(v)
\Lambda_{\nu\alpha}(v,u_p)(\hat q-\hat k)_\alpha\, ,\quad\quad
\tilde q^\mu=\frac{1}{2}\,\underline{\Lambda}^{\mu\nu}(v)
\Lambda_{\nu\alpha}(v,u_q)
(\hat p-\hat k)^\alpha\, ,
\nonumber\\[2mm]
k^\mu&=&K^\mu-p^\mu-q^\mu\, ,\quad\quad
s_p=(\hat q+\hat k)^2\, ,\quad\quad
s_q=(\hat p+\hat k)^2\, ,
\nonumber\\[2mm]
u_q^\mu&=&(\hat p+\hat k)^\mu/\sqrt{s_q}\, ,\quad\quad
u_p^\mu=(\hat q+\hat k)^\mu/\sqrt{s_p}\, ,\quad\quad
\en
and, for any four-vector $a^\mu$, one has
\eq
\hat a^\mu=a^\mu-v^\mu(av)+v^\mu w_v(a)\, .
\en
Furthermore,
\eq
\Delta&=&(K^2-9m^2)\, ,\quad\quad
\Delta_p=(K-\hat p)^2-4m^2\, ,\quad\quad
\Delta_q=(K-\hat q)^2-4m^2\, ,\quad\quad
\en
and the function $f_\ell(s)$ is related to the scattering phase, according to
Eq.~(\ref{eq:f2}).

\subsection{Relativistic invariance}

In order to establish the relativistic invariance of the above infinite-volume
framework,\footnote{It is clear that the notion of relativistic invariance
applies to the infinite-volume case only. In a finite volume, the invariance is broken by a box. Hence, in a finite volume, the statement boils down to
a frame-independence of the couplings extracted from the fit to the energy
levels, up to exponential corrections.} one has to perform an arbitrary Lorentz-boost on all external momenta: $p=p_\Omega\mapsto\Omega p$,
$q=q_\Omega\mapsto\Omega q$,
$K=K_\Omega\mapsto\Omega K$ and $v=v_\Omega\mapsto\Omega v$.
The integration variable undergoes the same boost
$k=k_\Omega\mapsto\Omega k$. All functions depending on Lorentz-scalars
are, of course, manifestly Lorentz-invariant. Hence, the question boils down
to the transformation of the three-vectors
$\underline{\bf p},\underline{\bf q}$ and ${\bf \tilde p},{\bf \tilde q}$.

Let us first consider the four-vector $\underline p=\underline{\Lambda}(v)\hat p$,
where  $\underline{\Lambda}(v)$ is defined from the condition
$\underline{\Lambda}(v)v=v_0$. In the boosted frame, one has
$\underline{\Lambda}(v_\Omega)v_\Omega=v_0$.
This gives
\eq
\underline{\Lambda}(v_\Omega)=R\underline{\Lambda}(v)\Omega^{-1}\, ,
\en
where
$R$ is a pure rotation that does not depend on the choice of the four-vector $p$.
If $\Omega$ is a pure
rotation itself, then $R=\Omega$.
Hence,
\eq
\underline{p}_\Omega=\underline{\Lambda}(v_\Omega)\hat p_\Omega
=R\underline{\Lambda}(v)\Omega^{-1}\Omega \hat p=R\underline{p}\, .
\en
The same line of reasoning holds for the four-vector
$\tilde p=\underline{\Lambda}(v)\Lambda(v,u)\hat p$, since the vector
$\Lambda(v,u)\hat p\mapsto \Omega \Lambda(v,u)\hat p$ transforms
exactly as the vector $\hat p$ (we remind the reader that $\Lambda(v,u)$, in difference
of $\underline{\Lambda}(v)$, is the Lorentz-tensor).
The transformation matrix $R=R(\Omega,v,u)$, which depends on the parameters of the $\Omega$,
as well as on the vectors $v$ and $u$ in a non-linear manner, is the same in both cases.\footnote{The rotation $R$ is related to the Thomas-Wigner rotation~\cite{Thomas:1926dy,Wigner:1939cj}.}

To summarize, it is seen that the Lorentz transformations, acting  
on the three-vectors ${\bf \tilde p},\underline{\bf p},\ldots$, result in the $SO(3)$
transformations whose parameters can be expressed through the parameters of the
initial Lorentz transformations as well as the vectors $v$ and $u$.
The transformation of the
various quantities that enter the kernel of the Faddeev equation can be defined
through the transformation properties of spherical functions, entering this expression:
\eq\label{eq:Wignerd}
\mathscr{Y}_{\ell m}(R{\bf p})=\sum_{m'}\left(\mathscr{D}^{(\ell)}_{mm'}(R)\right)^*
\mathscr{Y}_{\ell m'}({\bf p})\, ,
\en
where $\mathscr{D}^{(l)}_{mm'}(R)$ denote Wigner $D$-functions. Furthermore,
The kernel of the equation $Z=Z_{\sf ex}+Z_{\sf loc}$ consists of two parts, corresponding
to the exchange diagram and the local particle-dimer interaction, see the first and the second terms in Eq.~(\ref{eq:driving}), respectively. The transformation of the first term
is straightforwardly defined by Eq.~(\ref{eq:Wignerd}):
\eq
\left(Z_{\sf ex}\right)_{\ell'm',\ell m}(\Omega p,\Omega q)=
\sum_{m'''m''}\mathscr{D}^{(\ell')}_{m'm'''}(R)
\left(Z_{\sf ex}\right)_{\ell'm''',\ell m''}(p,q)
\left(\mathscr{D}^{(\ell)}_{mm''}(R)\right)^*\, .
\en
Establishing the transformation properties of the local term is a bit trickier and will be considered in Appendix~\ref{app:Zloc}. Here, we simply state that
$Z_{\sf loc}$ transforms exactly in the same way as  $Z_{\sf ex}$. Hence,
their sum has the same property. Furthermore, the propagator $S_\ell$ is invariant under
the Lorentz transformations. From this, one finally concludes that the particle-dimer
scattering amplitude, which is a solution of the Faddeev equation~(\ref{eq:Faddeev}), has
the same transformation property as the kernel $Z$:
\eq
T_{\ell'm',\ell m}(\Omega p,\Omega q)=
\sum_{m'''m''}\mathscr{D}^{(\ell')}_{m'm'''}(R)
T_{\ell'm''',\ell m''}(p,q)
\left(\mathscr{D}^{(\ell)}_{mm''}(R)\right)^*\, .
\en
This is nothing but the statement about the manifest Lorentz invariance of the framework.

\subsection{Faddeev equation in a finite volume and the quantization condition}\label{sec:QC}

In a finite volume, the integration over three-momenta is replaced
by the sums. In this course, the Faddeev equations, displayed above,
undergo some modifications. As in the infinite volume, we start from
the propagator of a dimer
\eq
i\langle 0|T[T_{\ell' m'}(x)T^\dagger_{\ell m}(y)]|0\rangle=
\int\frac{dP^0}{2\pi}\,\frac{1}{L^3}\sum_{\bf P}e^{-iP(x-y)}S^L_{\ell' m',\ell m}(P)\, .
\en
Here, the dimer three-momentum runs over the discrete values
${\bf P}=\dfrac{2\pi}{L}\,{\bf n}\, ,~{\bf n}\in \mathbb{Z}^3$. Note that,
owing to the lack of the rotational invariance in a finite volume, the
propagator is no more diagonal in the indices $\ell m$ and $\ell' m'$.
It obeys the Dyson-Schwinger equation
\eq
S^L_{\ell' m',\ell m }(P)=-\frac{1}{\sigma_{\ell'}}\,\delta_{\ell'\ell}\delta_{m'm}
-\frac{1}{\sigma_{\ell'}}\sum_{\ell''m''}\Sigma^L_{\ell' m',\ell''m''}(P)S^L_{\ell'' m'',\ell m }(P)\, ,
\en
where
\eq\label{eq:Sigma}
\Sigma^L_{\ell' m',\ell m}(P)=f_{\ell'}(P^2)f_\ell(P^2)
\int\frac{dq^0}{2\pi i}\,\frac{1}{2L^3}\sum_{\bf q}
\frac{\left(\mathscr{Y}_{\ell' m'}({\bf \tilde q})\right)^*
  \mathscr{Y}_{\ell m}({\bf\tilde q})}
     {(m^2-q^2-i\varepsilon)(m^2-(P-q)^2-i\varepsilon)}\, ,\quad\quad
     \en
     where
     \eq\label{eq:momenta}
     \tilde q^\mu=\frac{1}{2}\,\underline{\Lambda}^{\mu\nu}(v)
     \Lambda_{\nu\alpha}(v,u)(\hat q-\hat q')^\alpha\, ,
     \quad\quad
     u^\mu=\frac{(\hat q+\hat q')^\mu}{(\hat q+\hat q')^2}\, ,\quad\quad
     q'=P-q\, .
     \en
        Carrying out the integration over $q^0$ in Eq.~(\ref{eq:Sigma}), one obtains
 \eq\label{eq:Sigma1}
\Sigma^L_{\ell' m',\ell m }(P)=f_{\ell'}(P^2)f_\ell(P^2)
\frac{1}{2L^3}\sum_{\bf q}\frac{\left(\mathscr{Y}_{\ell' m'}({\bf \tilde q})\right)^*\mathscr{Y}_{\ell m }({\bf\tilde q})}
     {2w({\bf q})2w({\bf P}-{\bf q})(w({\bf q})+w({\bf P}-{\bf q})-P^0)}\, .\quad\quad
          \en       
          Note that the propagator still implicitly
          depends on $v^\mu$, since $v^\mu$ enters the definition
          of any on-mass-shell vector $\hat a^\mu$. This dependence comes however only from the numerator
          and can be worked out explicitly. We relegate this task to Appendix~\ref{app:dimer_prop} where, in particular, it will be shown how does
          one systematically factor out this dependence. Moreover, already at this stage it is
          seen that
the $v$-dependence disappears in the S-wave, as claimed in Ref.~\cite{Muller:2021uur}.

The Faddeev equation in a finite volume can be written as           
\eq\label{eq:Faddeev-finite}
T^L_{\ell'm',\ell m}(p,q)
&=&Z_{\ell'm',\ell m}(p,q)
+\sum_{\ell'''m''',\ell''m''}\frac{1}{L^3}\sum^{\Lambda_v}_{\bf k}\frac{1}{2w({\bf k})}\,
Z_{\ell'm',\ell''' m'''}(p,k)
\nonumber\\[2mm]
&\times&
S^L_{\ell'''m''',\ell''m''}(K-k)T^L_{\ell''m'',\ell m}(k,q)\, ,
\en
where
\eq
\sum^{\Lambda_v}_{\bf k}f(k)\doteq\sum_{\bf k}\theta(\Lambda^2+k^2-(kv)^2)f(k)\, .
\en
The three-body quantization condition takes the form
$\det \mathscr{A}=0$, where $\mathscr{A}$ is a matrix both in the spectator momenta $p,q$,
as well as the partial-wave indices $\ell m,\ell'm'$:
\eq\label{eq:A}
\mathscr{A}_{\ell' m',\ell m}(p,q)=2w({\bf p})\delta_{{\bf p}{\bf q}}\left(
  S^L_ {\ell' m',\ell m }(K-p)\right)^{-1}
-\frac{1}{L^3}\,Z_{\ell' m',\ell m}(p,q)\, .
\en
Here,
\eq
\left(S^L_ {\ell' m',\ell m}(K-p)\right)^{-1}=-\delta_{\ell'\ell}\delta_{m'm}\sigma_{\ell'}
-\Sigma^L_ {\ell' m',\ell m}(K-p)\, .
\en

\subsection{Reduction of the quantization condition}

Using symmetry under the octahedral group (or the little groups thereof),
one may achieve a partial diagonalization of the quantization condition.
Namely, let ${\cal G}$ be a subgroup of the octahedral group $O_h$ that leaves
the vector ${\bf K}$ invariant. Since $v^\mu$ is chosen to be parallel to
$K^\mu$, the vector ${\bf v}$ is invariant under ${\cal G}$ as well.
Hence, under the transformations from the group ${\cal G}$, the matrix $\mathscr{A}$
from Eq.~(\ref{eq:A}) transforms as
\eq
\mathscr{A}_{\ell' m',\ell m}(gp,gq)=\sum_{m'''m''}\mathscr{D}^{(\ell')}_{m'm'''}(g)
\mathscr{A}_{\ell' m''',\ell m''}(p,q)\left(\mathscr{D}^{(\ell)}_{mm''}(g)\right)^*\, ,
\quad\quad g\in {\cal G}\, .
\en
It is well known that the linear space, in which the irreducible representation (irrep) of the $SO(3)$ group with the angular momentum $\ell$ is realized, falls into different orthogonal subspaces, corresponding to the irreducible representations of the octahedral group or the little groups thereof. The basis vectors of the irreps of two groups are related by a linear transformation
\eq
\mathscr{Y}_{\lambda (t\Delta)}^\ell=\sum_m c^{\ell m}_{\lambda (t\Delta)}
\mathscr{Y}_{\ell m}\, .
\en
Here, $\Delta$ denotes an irrep of the group ${\cal G}$, $t$ labels different copies
of the same irrep $\Delta$, and $\lambda$ is an index, corresponding to different basis
vectors of a given irrep. The coefficients $c^{\ell m}_{\lambda t\Delta}$ are well known an,
for small values of $\ell$, are tabulated, e.g., in Ref.~\cite{Gockeler:2012yj}. These
coefficients obey the orthogonality conditions
\eq
\sum_m \left(c^{\ell m}_{\lambda' (t'\Delta')}\right)^* c^{\ell m}_{\lambda (t\Delta)}
&=&\delta_{t't}\delta_{\Delta'\Delta}\delta_{\lambda'\lambda}\, ,
\nonumber\\[2mm]
\sum_{t\Delta\lambda} \left(c^{\ell m'}_{\lambda (t\Delta)}\right)^* c^{\ell m}_{\lambda (t\Delta)}
&=&\delta_{mm'}\, .
\en
Besides this, we shall need to define the Clebsch-Gordan coefficients for the group ${\cal G}$. Note that the octahedral group $O_h$ as well as the little groups $C_{4v},C_{2v},C_{3v}$, corresponding to a different choice of the center-of-mass momentum, are simply reducible. Since all the representations are unitary, these Clebsch-Gordan coefficients can be chosen to be real. The orthogonality condition for the Clebsch-Gordan coefficients takes the form
\eq
\sum_{\Gamma\alpha}\langle \Sigma\rho,\Delta\lambda|\Gamma\alpha\rangle
\langle \Sigma\rho',\Delta\lambda'|\Gamma\alpha\rangle&=&\delta_{\rho\rho'}\delta_{\lambda\lambda'}\, ,
\nonumber\\[2mm]
\sum_{\rho\lambda}\langle \Sigma\rho,\Delta\lambda|\Gamma\alpha\rangle
\langle \Sigma\rho,\Delta\lambda|\Gamma'\alpha'\rangle&=&\delta_{\Gamma\Gamma'}\delta_{\alpha\alpha'}\, ,
\en
where capital and small Greek letters label the irreps and the basis vectors in a given irrep, respectively.

Defining now
\eq
\mathscr{A}_{\lambda'(t'\Delta'),\lambda(t\Delta)}^{\ell'\ell}(p,q)=\sum_{m'm}
\left(c^{\ell'm'}_{\lambda'(t'\delta')}\right)^*
\mathscr{A}_{\ell'm',\ell m}(p,q)
c^{\ell m}_{\lambda(t\Delta)}\, ,
\en
it is easy to show that this quantity transforms as
\eq\label{eq:transformation}
\mathscr{A}_{\lambda'(t'\Delta'),\lambda(t\Delta)}^{\ell'\ell}(gp,gq)
=\sum_{\lambda'''\lambda''}T^{(\Delta')}_{\lambda'\lambda'''}(g)
\mathscr{A}_{\lambda'''(t'\Delta'),\lambda''(t\Delta)}^{\ell'\ell}(p,q)
T^{(\Delta)}_{\lambda''\lambda}(g^{-1})\, .
\en
Here, $T^{(\Delta)}(g)$ denotes the matrix of an irreducible representation $\Delta$.

At the next step, we define a projection
\eq\label{eq:projection}
\mathscr{A}_{\sigma'(t'\Delta')\Sigma',\sigma(t\Delta)\Sigma}^{\ell'\Gamma'\alpha',\ell\Gamma\alpha}(p_r,q_s)
&=&\frac{s_{\sigma'}}{G}\,\frac{s_\Sigma}{G}\,
\sum_{g',g\in{\cal G}}\sum_{\lambda'\rho',\lambda\rho}
\langle\Sigma'\rho',\Delta'\lambda'|\Gamma'\alpha'\rangle T^{(\Sigma')}_{\rho'\sigma'}(g')
\nonumber\\[2mm]
&\times&\mathscr{A}_{\lambda'(t'\Delta'),\lambda(t\Delta)}^{\ell'\ell}(g'p_r,gq_s)
\langle\Sigma\rho,\Delta\lambda|\Gamma\alpha\rangle \left(T^{(\Sigma)}_{\rho\sigma}(g)\right)^*\, .
  \en
  Here, $p_r$ and $q_s$ denote reference momenta in the shells $r$ and $s$, respectively,
  $s_{\Sigma},s_{\Sigma'}$ are the dimensions of the pertinent irreps, and $G$ is the total
  number of the elements in the group ${\cal G}$
  (for a detailed discussion, see, e.g., Ref.~\cite{Doring:2018xxx}).
  It can be seen (see Appendix~\ref{app:cubic})
  that this matrix is diagonal in the irreps $\Gamma,\Gamma'$:
  \eq\label{eq:diagonal}
  \mathscr{A}_{\sigma'(t'\Delta')\Sigma',\sigma(t\Delta)\Sigma}^{\ell'\Gamma'\alpha',\ell\Gamma\alpha}(p_r,q_s)
  =\frac{s_{\Sigma'}s_\Sigma}{Gs_\Gamma}\,\delta_{\Gamma'\Gamma}\delta_{\alpha'\alpha} \mathscr{A}_{\sigma'(t'\Delta')\Sigma',\sigma(t\Delta)\Sigma}^{\ell'\ell;\Gamma}
  (p_r,q_s)\, ,
  \en
  where
  \eq\label{eq:AGamma}
  \mathscr{A}_{\sigma'(t'\Delta')\Sigma',\sigma(t\Delta)\Sigma}^{\ell'\ell;\Gamma}
  (p_r,q_s)&=&\sum_{g\in {\cal G}}\sum_{\lambda'\lambda}\sum_{\sigma''\gamma}
  \langle \Sigma'\sigma'',\Delta'\lambda'|\Gamma\gamma\rangle
  \langle\Sigma\sigma,\Delta\lambda|\Gamma\gamma\rangle
\nonumber\\[2mm]
  &\times&T^{(\Sigma')}_{\sigma''\sigma'}(g)\mathscr{A}_{\lambda'(t'\Delta'),\lambda(t\Delta)}^{\ell'\ell}(gp_r,q_s)\, .
  \en
  This means that the quantization condition $\det\mathscr{A}=0$ diagonalizes
  in different irreps $\Gamma$, taking the form $\det\mathscr{A}^\Gamma=0$,
  where the matrix $\mathscr{A}^\Gamma$ is defined by Eq.~(\ref{eq:AGamma}).

\subsection{Three-particle decays}

In the following, for brevity, we shall refer to the heavy and light particles
to as ``kaons'' and ``pions,'' respectively, bearing in mind the weak decays of kaons
into three pions (of course, this analogy is not full since it does not take into account
the isospin.).
The Lagrangian that describes the decay of the ``kaon'' into a ``pion''-dimer
pair can be written as
\eq\label{eq:LG}
\mathscr{L}_G=
\sqrt{4\pi}\sum_{\ell m}
\frac{(-1)^\ell}{\sqrt{2\ell+1}}\,
\left(K^\dagger
G_\ell(\Delta_T)
\left((\mathscr{Y}_{\ell, -m}(\underline{\bf w}))^*\phi\right)T_{\ell m}
+\mbox{h.c.}\right)\, .
\en
Here, $G_\ell$ is a low-energy polynomial
\eq\label{eq:LK}
G_\ell(\Delta_T)
=G^{(\ell)}_0+G^{(\ell)}_1\Delta_T+\cdots\, .
\en
Here, $\Delta_T$ is defined in Eq.~(\ref{eq:DeltaT}). The coefficients of the Taylor expansion $G^{(\ell)}_0,G^{(\ell)}_1,\ldots$ play the role of the effective
couplings, and $\Delta_T$ is of order $p^2$ in the NREFT power counting.
It is assumed that $G^{(\ell)}_0,G^{(\ell)}_1,\ldots$ are proportional to an intrinsically
small coupling (an analog of the Fermi coupling in weak kaon decays). All terms in the
decay amplitude at the second and higher orders in these couplings are neglected.
Furthermore, integrating out the dimer field, one arrives at the Lagrangian
that describes decays of ``kaons'' into three ``pions'' at tree level, analogous
to the one displayed in Refs.~\cite{Colangelo:2006va,Gasser:2011ju}. The Bose-symmetry
in the final state imposes constraints on the couplings $G^{(\ell)}_0,G^{(\ell)}_1,\ldots$.
An exact number of these constraints however depends on the details of the final-state
interactions, see the discussion in Sect.~\ref{sec:couplings}.\footnote{The expansion Eq.~(\ref{eq:LK}) is an analog of the expansion of the quantity
$A_{K3\pi}^{\textrm{PV}}$ from Ref.~\cite{Hansen:2021ofl} in the particle-dimer formalism. The constraints imposed by the Bose-symmetry are already taken into account in that paper.}

This Lagrangian should be supplemented by the Lagrangian describing the free ``kaon''
\eq
\mathscr{L}_K=K^\dagger2w_v^K(iv\partial-w^K_v)K\, .
\en
Here, $w_v^K=\sqrt{M^2+\partial^2-(v\partial)^2}$ and $M$ denotes the mass of
the ``kaon''.
The full Lagrangian of the system is given by $\mathscr{L}+\mathscr{L}_G+\mathscr{L}_K$, where the individual terms are given by Eqs.~(\ref{eq:L}), (\ref{eq:LG}) and (\ref{eq:LK}), respectively.

\subsection{Convergence of the NREFT approach}

The convergence of the NREFT approach is a subtle issue which comprises
several distinct  aspects of the problem in question. Below, we wish to put this
issue under scrutiny. To start with, a rule of thumb requiring $p/m\ll 1$,
where $p$ denotes the magnitude of a characteristic three-momentum in the process
seems too restrictive, say, in kaon decays. From the point of view of kinematics,
this restriction is not relevant: all low energy singularities in the amplitudes
appear exactly at the right place~\cite{Colangelo:2006va,Gasser:2011ju}. So, everything
finally boils down to a question, whether the real part of the decay amplitude can be well
approximated by a low-energy polynomial. Were this not the case, one would conclude
that the contributions which are not explicitly taken into account in NREFT
(say, all kinds of the $t$- and $u$-channel diagrams, multi-particle intermediate
states and so on), play an important role and NREFT is not applicable. Fortunately,
one can check this assumption experimentally in the real kaon decays, since the
decay amplitudes are measured very accurately. The experiment neatly confirms
the conjecture: the real part of the amplitude can be indeed fitted by a polynomial
in the Mandelstam variables $s_i$ of a rather low order in the whole region of the Dalitz
plot (see, e.g.~\cite{Batley:2009ubw}).
This justifies the use of the NREFT approach for the description of the
kaon decays into three pions, despite the fact that characteristic momenta of pions
are of order of the pion mass itself.

Two remarks are in order. The first one concerns the choice of the expansion point.
The ``standard'' version of the NREFT implies the expansion of the amplitudes in the
external three-momenta. For example, in case of the two-particle scattering, this corresponds
to the effective-range expansion which is performed around threshold. The convergence
of such an expansion in some cases
might be affected by the presence of the subthreshold singularities
in the initial relativistic theory (for instance, effective-range expansion in the vicinity
of the $\rho$-meson fails due to the left-hand cut in the $\pi\pi$ amplitude). In such cases,
choosing the expansion point somewhere above threshold allows one to circumvent the
problem, if the effect of the inelastic channels is still small. The NREFT framework is flexible
enough and can be adapted to the change of the expansion point with a minor
modification only. We do not display a modified Lagrangian here, because this issue
is not directly related to the main problem considered in the present paper.

The second remark is general. One may ask, whether NREFT is applicable to study of the
generic three-particle decays beyond $K\to 3\pi$.
There exists no unique answer to this (very difficult) question.
It should be therefore should be addressed on the case-to-case basis:
for example, it would be conceivable that NREFT is still valid, if
it is {\it a priori} known that the inelastic channels are not important for a process in question.

\section{Derivation of the three-particle LL formula in the relativistic-invariant framework}

\label{sec:derivation}

\subsection{The Wick rotation}

In the derivation of the 3-particle LL formula we closely follow the path outlined
in Ref.~\cite{Muller:2020wjo}. There are, however, some differences caused by an
explicit Lorentz invariance. These differences will be highlighted in the following.

We start with discussing the Wick rotation in case of a quantization axis chosen
parallel to an arbitrary timelike unit vector $v^\mu$. To be specific, we consider the
simplest possible choice of the three-particle source/sink operator
$\mathscr{O}(x)=\phi^3(x)$. As the first step in the derivation, we would like to extract
the finite-volume matrix element $\langle 0|\mathscr{O}|n\rangle$, where $|n\rangle$
denotes a (discrete) eigenstate of the total Hamiltonian in a finite volume. To this end,
we consider the two-point function
\eq
G(x-y)=\langle 0|T\mathscr{O}(x)\mathscr{O}^\dagger(y)|0\rangle\, .
\en
Using the translational invariance and the closure relation, in the Minkowski space we get
\eq\label{eq:thesame}
G(x-y)=\sum_ne^{-iP_n(x-y)}\left|\langle 0|\mathscr{O}(0)|n\rangle\right|^2\, .
\en
The three-momentum of the intermediate state is quantized, 
${\bf P}_n=\dfrac{2\pi}{L}\,{\bf n},~{\bf n}\in\mathbb{Z}^3$. The time component
of the four vector is given by $P_n^0$, where it is assumed that $P_n^0> |{\bf P}|$.
 In order to {\em define} the Wick rotation in the frame moving with the four-velocity
 $v^\mu$, one defines the parallel and transverse components:
 \eq
 (x-y)_\parallel&=&v(x-y)\, ,\quad\quad
 (x-y)_\perp^\mu=(x-y)^\mu-v^\mu(x-y)_\parallel\, ,
\nonumber\\[2mm]
 P_{n,\parallel}&=&vP_n\, ,\quad\quad
 P_{n,\perp}^\mu=P_n^\mu-v^\mu P_{n,\parallel}\, ,
 \en
 so that
 \eq
 P^\mu(x-y)_\mu=P_{n,\parallel} (x-y)_\parallel
 +P_{n,\perp}^\mu (x-y)_{\perp,\mu}\, .
 \en
 The Wick rotation in the moving frame is defined through the analytic continuation
 \eq
 (x-y)_\parallel\to -i(x-y)_\parallel\, ,\quad\quad (x-y)_\perp^\mu\to (x-y)_\perp^\mu\, .
 \en
 Furthermore, since $P_n^0>|{\bf P}_n|$, then
 \eq
 P_{n,\parallel}=v^0P_n^0-{\bf v}{\bf P}_n\geq v^0P_n^0-|{\bf v}|\,|{\bf P}_n|>0\, .
 \en
 Thus, in the limit $(x-y)_\parallel\to\infty$, all exponentials are damping
 in the Euclidean space and one gets
 \eq\label{eq:damping}
G(x-y)=\sum_ne^{-P_{n,\parallel}(x-y)_\parallel-iP_{n,\perp}(x-y)_\perp}\left|\langle 0|\mathscr{O}(0)|n\rangle\right|^2\, ,
\en
where {\em the same matrix elements} as in Eq.~(\ref{eq:thesame}) appear in the right-hand side.
It should be noted however that, on the lattice, one does not need calculate the
Green functions, in which the continuation to the Euclidean space is performed in
moving frames. This is a purely theoretical construction that will merely help us to derive
3-particle LL formula in all frames.

\subsection{The matrix element}

\begin{figure}[t]
  \begin{center}
    \includegraphics*[width=9.cm]{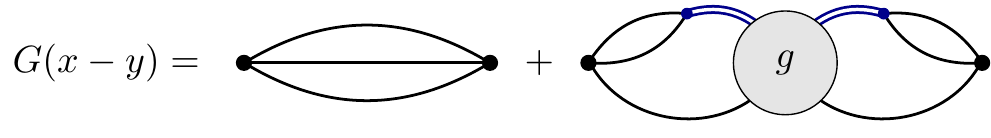}
    
    \vspace{0.5cm}
    
    \includegraphics*[width=9.cm]{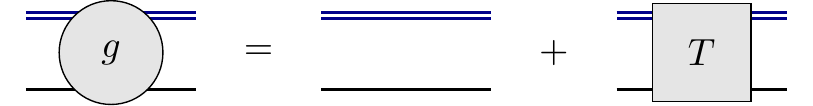}
    \caption{Two-point function, calculated in the effective field theory. The solid and double lines stand for the one-particle propagator and the full dimer propagator, respectively.
    The quantity $T$ denotes the particle-dimer scattering amplitude. }
    \label{fig:twopoint-3-LL}
  \end{center}
\end{figure}

Next, let us evaluate the two-point function $G(x-y)$ in NREFT, starting from the infinite volume. It will be useful to use the components $(a_\parallel,a_\perp^\mu)$
instead of $(a^0,{\bf a})$ for all four vectors $a^\mu$. For simplicity, we shall work in the Minkowski space and perform the Wick rotation only at the very end. The Feynman diagrams, contributing to the above two-point function, can be summed up, resulting in a compact expression
\eq\label{eq:compact}
G(x-y)=3^2\int\frac{dK_\parallel}{2\pi i}\,\frac{d^3K_\perp}{(2\pi)^3}\,
e^{-iK_\parallel(x-y)_\parallel-iK_\perp(x-y)_\perp}G(K)\, ,
\en
where $G(k)=G_0(K)+G_1(k)$ and
\eq\label{eq:G0G1}
G_0(K)&=&\frac{2}{3}\,\int\frac{d^3 q_{1,\perp}}{(2\pi)^3 2w_v(q_1)}\,
\frac{d^3 q_{2,\perp}}{(2\pi)^3 2w_v(q_2)}\,
\frac{1}{2w_v(K-q_1-q_2)}
\nonumber\\[2mm]
&\times&\frac{1}{\left(w_v(q_1)+w_v(q_2)+w_v(K-q_1-q_2)-K_\parallel-i\varepsilon\right)}\, ,
\nonumber\\[2mm]
G_1(K)&=&4\pi\sum_{\ell m,\ell'm'}
\int\frac{d^3 q_{1,\perp}}{(2\pi)^3 2w_v(q_1)}\,
\frac{d^3 q_{2,\perp}}{(2\pi)^3 2w_v(q_2)}\,
\frac{d^3 k_{1,\perp}}{(2\pi)^3 2w_v(k_1)}\,
\frac{d^3 k_{2,\perp}}{(2\pi)^3 2w_v(k_2)}
\nonumber\\[2mm]
&\times&
\frac{\mathscr{Y}_{\ell' m'}({\bf \tilde q}_1)f_{\ell'}((K-k_1)^2)}
{2w_v(K-q_1-k_1)
  \left(w_v(q_1)+w_v(k_1)+w_v(K-q_1-k_1)-K_\parallel-i\varepsilon\right)}\,
\nonumber\\[2mm]
&\times&g_{\ell' m',\ell m }(k_1,k_2;K)
\nonumber\\[2mm]
&\times&\frac{f_\ell((K-k_2)^2)\left(\mathscr{Y}_{\ell m}({\bf \tilde q}_2)\right)^*}
{2w_v(K-q_2-k_2)
  \left(w_v(q_2)+w_v(k_2)+w_v(K-q_2-k_2)-K_\parallel-i\varepsilon\right)}\, .
\en
Furthermore,
\eq\label{eq:smallg}
g_{\ell' m',\ell m}(k_1,k_2;K)&=&
2w_v(k_1)\delta^3(k_{1,\perp}-k_{2,\perp})\delta_{\ell\ell'}\delta_{mm'}S_\ell(K-k_1)
\nonumber\\[2mm]
  &+&S_{\ell'}(K-k_1)T_{\ell' m',\ell m}(k_1;k_2)S_\ell(K-k_2)\, .
  \en
Here, 
${\bf \tilde q}_1$ and ${\bf \tilde q}_2$ are relative momenta of particle pairs,
boosted to the CM frames of pertinent dimers. Graphically, the sum of all diagrams
is displayed in Fig.~\ref{fig:twopoint-3-LL}.

The spectral representation of the particle-dimer scattering amplitude
$T_{\ell' m',\ell m}(k_1,k_2)$ (again, in the infinite volume), can be written as follows
\eq\label{eq:spectral}
T_{\ell' m',\ell m}(k_1,k_2)=\sum_{n}\
\frac{\psi_{n}^{(\ell' m')}(k_1)
  \bar \psi_{n}^{(\ell m)}(k_2)}
{P_{n,\parallel}-K_\parallel-i\varepsilon}\, .
\en
Here, the index $n$ labels the eigenvalues of the Hamiltonian in a given channel, and
$\psi_{n}^{(\ell' m')}(k_1)$,  $\bar \psi_{n}^{(\ell m)}(k_2)$ stand for the wave
function and its conjugate, respectively.

In order to write down a finite-volume counterpart of the above equations, one has first
to boost the integration momenta to the rest frame (because the discretization is done
always in the rest frame). This can be achieved by using
\eq\label{eq:boost_measure}
\frac{d^3q_{i,\perp}}{(2\pi)^32w_v(q_i)}=\frac{d^3{\bf q}_i}{(2\pi)^32w({\bf k})}\, ,
\quad\quad\mbox{and similarly for $k_i$\, ,}
\en
while
\begin{equation}
	\frac{d^3K_{\perp}}{(2\pi)^3 2\sqrt{P_n^2 - K_{\perp}^2}} = \frac{d^3{\bf K}}{(2\pi)^3 2\sqrt{P_n^2 + {\bf K}^2}} \,.
      \end{equation}
Note also that one has adjusted the volume of the box so that $P_n^0=\sqrt{M_K^2+{\bf P}_n^2}$. Consequently, the four-momentum of the state $|n\rangle$ transforms similarly to the four-momentum of a single particle that is, generally, not the case for the multi-particle states in a finite volume.

Writing down Eq.~(\ref{eq:G0G1}) in a finite volume, using Eqs.~(\ref{eq:smallg}) and (\ref{eq:spectral}), and
carrying out the contour integration over $K_\parallel$, one arrives at
\eq\label{eq:pole}
G(x-y)&=&3^2\sum_{\ell m,\ell'm',\ell''m'',\ell'''m'''}
\frac{4\pi}{L^{15}}
\nonumber\\[2mm]
&\times&\sum_{{\bf q}_1{\bf q}_2{\bf k}_1{\bf k}_2}
\frac{e^{-iP_{n,\parallel}(x-y)_\parallel-iK_\perp (x-y)_\perp}
	\sqrt{M_K^2-K_\perp^2}}
{2w({\bf q}_1)2w({\bf q}_2)2w({\bf k}_1)2w({\bf k}_2)\sqrt{M_K^2 + {\bf K}^2}}
\nonumber\\[2mm]
&\times&\frac{\mathscr{Y}_{\ell' m'}({\bf \tilde q}_1)
	f_{\ell'}((K-k_1)^2)S^L_{\ell' m',\ell'''m'''}(K-k_1)
	\psi_n^{(\ell'''m''')}(k_1)}
{2w_v(K-q_1-k_1)
	\left(w_v(q_1)+w_v(k_1)+w_v(K-q_1-k_1)-P_{n,\parallel}\right)}\,
\nonumber\\[2mm]
&\times&\frac{\bar\psi^{(\ell''m'')}(k_2)S^L_{\ell''m'',\ell m}(K-k_2)
	f_\ell((K-k_2)^2)\left(\mathscr{Y}_{\ell m}({\bf \tilde q}_2)\right)^*}
{2w_v(K-q_2-k_2)
	\left(w_v(q_2)+w_v(k_2)+w_v(K-q_2-k_2)-P_{n,\parallel}\right)} + \dots \, ,\quad\quad
\en
where
\eq\label{eq:Kperp_Ppar}
K^0_\perp=-\frac{{\bf v}^2}{v^0}\,P_{n,\parallel}
+\frac{{\bf v}{\bf K}}{v^0}\, ,\quad\quad
{\bf K}_\perp={\bf K}-{\bf v}P_{n,\parallel}\, .
\en
In the equation above, we have explicitly singled out the contribution from a pole that emerges at $K_\parallel=P_{n,\parallel}$ with $P_n^2 = M_K^2$ for a given total three-momentum $\mathbf{K}$ (or, equivalently, for a fixed ${\bf K}_\perp$, as seen from Eq.~(\ref{eq:Kperp_Ppar})). The ellipsis stands for other single-pole contributions, $K_\parallel=P_{n,\parallel}$, emerging in the spectral decomposition of particle-dimer amplitude.
The individual amplitudes feature many more singularities in this variable. However, we also take into account the fact that, in the rest frame, all these singularities cancel. This property should persist in moving frames as well, since the position of the poles in our framework is Lorentz-invariant.

Now, continuing to the Euclidean space and comparing Eq.~(\ref{eq:damping})
with Eq.~(\ref{eq:pole}), one may finally read off
the matrix element one is looking for:
\eq\label{eq:matrix_element}
\left|\langle 0|\mathscr{O}(0)|n\rangle\right|&=&\frac{3\sqrt{4\pi}}{L^{3/2}}
\left(\frac{M_K}{\sqrt{M_K^2+{\bf K}^2}}\right)^{1/2}\,\biggl|\sum_{\ell m,\ell'm'}
\frac{1}{L^6}\sum^{\Lambda_v}_{{\bf q}{\bf k}}
\frac{\mathscr{Y}_{\ell' m'}({\bf \tilde q})}
{2w({\bf q})2w({\bf k})}\,
\nonumber\\[2mm]
&\times&\frac{f_{\ell'}((K-k)^2)S^L_{\ell' m',\ell m}(K-k)
\psi_n^{(\ell m )}(k)}
{2w_v(K-q-k)
  \left(w_v(q)+w_v(k)+w_v(K-q-k)-P_{n,\parallel}\right)}\biggr|\, .
\en
Here a cutoff on the momenta, which has been implicit in all previous
expressions, is displayed.
We also remind the reader that $|n\rangle$ denotes an eigenstate
with the momentum ${\bf K}$ (no summation over ${\bf K}$). The components of
the four-vector $K^\mu$ are:
\eq
K^\mu=\left(\frac{1}{v^0}\,(P_{n,\parallel}+{\bf v}{\bf K}),{\bf K}\right)\, .
\en
Furthermore, since there is no more summation in ${\bf K}$, we may fix the vector $v^\mu$ along $K^\mu$ that leads to $K^\mu_\perp=0$.
Finally, in a finite volume, $|n\rangle$ is a basis vector of some irreducible representation ${\it\Gamma}$ of some little group, or of an octahedral group (in case of ${\bf K}=0$).
Particular values of $\ell,\ell'$ contribute to this sum, if and only if ${\it\Gamma}$
is contained in the pertinent irreducible representation of the rotation group.

\subsection{Faddeev equation for the wave function}

The wave function, which was introduced in Eq.~(\ref{eq:spectral}), obeys the homogeneous Faddeev equation both in the infinite and in a finite volume, which can be straightforwardly obtained by substituting the ansatz defined by Eq.~(\ref{eq:spectral}) into the equation
for the particle-dimer scattering amplitude~(\ref{eq:Faddeev-finite})
and identifying pole terms on the both sides
of this equation. In a finite volume, this equation takes the form
\eq
\psi^{(\ell' m')}_n(p)&=&\sum_{\ell m,\ell''m''}
\frac{1}{L^3}\,\sum^{\Lambda_v}_{\bf k}\frac{1}{2w({\bf k})}\,
\nonumber\\[2mm]
&\times&Z_{\ell' m',\ell''m''}(p,k)S^L_{\ell''m'',\ell m}(K-k)
\psi^{(\ell m)}_n(k)\, .
\en
The infinite-volume analog of the above equation can be written as
\eq
\psi^{(\ell' m')}_n(p)&=&\sum_{\ell m }
\int^{\Lambda_v}\frac{d^3k_\perp}{(2\pi)^32w_v(k)}\,
Z_{\ell' m',\ell m}(p,k)S_\ell(K-k)
\psi^{(\ell m)}_n(k)\, .
\en
Since these are inhomogeneous equations, the normalization of the wave function should
be fixed independently. Using the standard technique (see, e.g., \cite{agrawala,Konig:2013wmc}), in a finite volume one obtains
\eq
1&=&\sum_{\ell' m',\ell m}\frac{1}{L^3}\,\sum^{\Lambda_v}_{\bf p}\frac{1}{2w({\bf p})}\,
\bar\psi^{(\ell' m')}_n(p)\left(\frac{d}{dK_\parallel}S^L_{\ell' m',\ell m}(K-p)\right)
\psi^{(\ell m)}_n(p)
\biggr|_{K_\parallel=P_{n,\parallel}}
\nonumber\\[2mm]
&+&\sum_{\ell m,\ell'm',\ell''m'',\ell'''m'''}
  \frac{1}{L^6}\,\sum^{\Lambda_v}_{{\bf p},{\bf q}}
\frac{1}{2w({\bf p})2w({\bf q})}\,
\bar\psi^{(\ell' m')}_n(p)S^L_{\ell' m',\ell'''m'''}(K-p)
\nonumber\\[2mm]
&\times&
\left(\frac{d}{dK_\parallel}Z_{\ell''' m''',\ell''m''}(p,q)\right)
S^L_{\ell'' m'',\ell m}(K-q)\psi^{(\ell m)}_n(q)
\biggr|_{K_\parallel=P_{n,\parallel}}\, .
\en
The infinite-volume counterpart of the above equation is
\eq
1&=&\sum_{\ell m}\int^{\Lambda_v}
\frac{d^3p_\perp}{(2\pi)^32w_v(p)}\,
\bar\psi^{(\ell m)}_n(p)\left(\frac{d}{dK_\parallel}S_\ell(K-p)\right)\psi^{(\ell m)}_n(p)
\biggr|_{K_\parallel=P_{n,\parallel}}
\nonumber\\[2mm]
&+&\sum_{\ell' m',\ell m}
\int^{\Lambda_v}\frac{d^3p_\perp}{(2\pi)^32w_v(p)}\,
\frac{d^3q_\perp}{(2\pi)^32w_v(q)}\,
\bar\psi^{(\ell' m')}_n(p)S_{\ell'}(K-p)
\nonumber\\[2mm]
&\times&
\left(\frac{d}{dK_\parallel}Z_{\ell' m',\ell m}(p,q)\right)
S_\ell (K-q)\psi^{(\ell m)}_n(q)
\biggr|_{K_\parallel=P_{n,\parallel}}\, .
\en

\subsection{The decay matrix element}

Let us now consider the two-point function $G_K(x)=\langle 0|T\mathscr{O}(x)J_K^\dagger(0)|0\rangle$,
where $J^\dagger_K(y)=\dfrac{\delta\mathscr{L}_G}{\delta K(y)}$ is a source operator
for the field $K(y)$. 

\begin{figure}[t]
  \begin{center}
    \includegraphics*{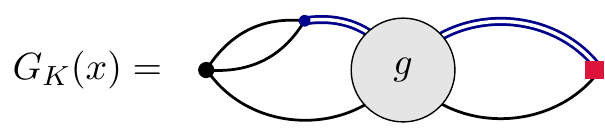}
  \end{center}
  \caption{Contributions to the two-point function $G_K(x)$. The single and double lines denote the particle propagator and the full dimer propagator, respectively. The quantity $T$
    stands for the particle-dimer scattering amplitude. The difference between $G_K(x)$ and
    the two-point function $G(x-y)$, shown in Fig.~\ref{fig:twopoint-3-LL}, boils down to the different sink operators used (these correspond to the utmost right vertices in the pertinent
    diagrams). }
  \label{fig:GK}
  \end{figure}

Applying the Feynman rules of the NREFT, in the infinite volume, the two-point function, shown in Fig.~\ref{fig:GK},  can be expressed as
\begin{equation}
	G_K(x) = 3 \int\frac{dK_\parallel}{2\pi i}\,\frac{d^3K_\perp}{(2\pi)^3}\,
	e^{-iK_\parallel x_\parallel-iK_\perp x_\perp} G_K(K) \,,
\end{equation}
where
\begin{align}
	G_K(K) &= 4\pi \sum_{\ell m,\ell'm'}   \frac{(-1)^\ell}{\sqrt{2\ell + 1}} \int\frac{d^3q_{1,\perp}}{(2\pi)^3 2w_v(q_1)}\,\frac{d^3k_{1,\perp}}{(2\pi)^3 2w_v(k_1)}\,\frac{d^3k_{2,\perp}}{(2\pi)^3 2w_v(k_2)} \nonumber\\[2mm]
	&\times \frac{f_{\ell'}((K-k_1)^2) \mathscr{Y}_{\ell'm'}(\mathbf{\tilde k}_1) g_{\ell'm', \ell m}(k_1,k_2;K) \mathscr{Y}_{\ell, -m}(\underline{\bf k}_2) G_\ell((K-k_2)^2)}{2w_v(K-q_1-k_1) \left(w_v(q_1) + w_v(k_1) + w_v(K-q_1-k_1) - K_\parallel - i \varepsilon \right)} \,,
\end{align}
with $G_\ell$ defined by Eq.~(\ref{eq:LK}).
Again, the corresponding expression in the finite-volume is obtained by boosting the integration momenta into the rest frame of the box, according to Eq.~\eqref{eq:boost_measure}. Using the spectral decomposition of the particle-dimer amplitude and carrying out the $K_\parallel$ integration:
\begin{align}\label{eq:GK_NREFT}
G_K(x) &= 3\,\sum_{\ell m, \ell'm', \ell''m'', \ell'''m'''} \frac{4\pi}{L^{12}}\frac{(-1)^\ell}{\sqrt{2\ell + 1}} \sum_{{\bf q}_1 {\bf k}_1{\bf k}_2} \frac{e^{-iP_{n,\parallel}x_\parallel}M_K}
{2w({\bf q}_1)2w({\bf k}_1)2w({\bf k}_2) \sqrt{M_K^2 + {\bf K}^2}} \nonumber\\[2mm]
&\times \frac{f_{\ell'}((K-k_1)^2) \mathscr{Y}_{\ell'm'}(\mathbf{\tilde k}_1) S^L_{\ell' m', \ell''' m'''}(K-k_1) \psi_{n}^{(\ell''' m''')}(k_1)}{2w_v(K-q_1-k_1) \left(w_v(q_1) + w_v(k_1) + w_v(K-q_1-k_1) - P_{n,\parallel} \right)} \nonumber\\[2mm]
&\times \bar \psi_{n}^{(\ell'' m'')}(k_2) S^L_{\ell'' m'', \ell m}(K-k_2) \mathscr{Y}_{\ell, -m}(\underline{\bf k}_2) G_\ell((K-k_2)^2) + \dots \,.
\end{align}
Similar to Eq.~\eqref{eq:pole}, we only display the contribution of the pole that emerges at $K_\parallel = P_{n,\parallel}$ with $P_n^2 = M_K^2$ for a given three-momentum $\mathbf{K}$.

On the other hand, by inserting a full set of states, we obtain
\begin{equation}
		G_K(x) = \sum_n e^{-iP_{n,\parallel}x_\parallel-iP_{n,\perp}x_\perp} \langle 0|\mathscr{O}(0)|n\rangle \langle n | J^\dagger_K(0) | 0 \rangle \, \,.
\end{equation}
Comparing this expression for $G_K$ with Eq.~\eqref{eq:GK_NREFT} and using the form of the matrix element in Eq.~\eqref{eq:matrix_element}, we can read off 
\begin{align}\label{eq:FV_decay_element} 
  L^{3/2} \langle n | J^\dagger_K(0) | 0 \rangle  &= \pm \sqrt{4\pi}
   \left(\frac{M_K}{\sqrt{M_K^2+{\bf K}^2}}\right)^{1/2}                                             \sum_{\ell m,\ell'm'} \frac{(-1)^\ell}{\sqrt{2\ell + 1}} \nonumber \\[2mm]
	&\times\frac{1}{L^3}\sum_{{\bf k}}^{\Lambda_v} \frac{1}{2w({\bf k})} \bar \psi_{n}^{(\ell' m')}(k) S^L_{\ell' m', \ell m}(K-k) \mathscr{Y}_{\ell, -m}(\underline{\bf k}) G_\ell((K-k)^2) \,.
\end{align}
Note that, due to the fact that the finite volume decay matrix element is real-valued, its phase is merely given by an undetermined sign. The choice of this sign is a delicate issue and is discussed in detail, say, in Ref.~\cite{Hansen:2021ofl}. In brief, as argued in that paper,
the phase of the eigenvector $|n\rangle$ is arbitrary, and one can always choose it
in a way that the sign of the (real-valued) matrix element
$\langle 0|\mathscr{O}(0)|n\rangle$ is positive.
In the following we shall stick to this convention,
which amounts to choosing ``+'' sign in the above equation.\footnote{In case of three weakly interacting particles in the final state, such a choice is a natural one, since
  the matrix element $\langle 0|\mathscr{O}(0)|n\rangle$, calculated in perturbation theory,
  starts with $3!\langle 0|\phi(0)|k\rangle^3$, where $|k\rangle$ denotes a
  state containing one
  free particle which is moving with a momentum $k$. Furthermore, the sign of the one-particle matrix element does not depend on $L$ and $k$. Thus, if the perturbative corrections are small and do not exceed the leading term, the set of matrix elements $\langle 0|\mathscr{O}(0)|n\rangle$, corresponding to the different $L$, $n$ and the total three-momentum ${\bf K}$, will have a positive relative sign. Note also that in the case of strong final-state interactions (i.e., when the two-particle resonances are present), the above argument is not {\it a priori}
  applicable and the question requires further scrutiny. We choose, however, not to elaborate further on this issue.
}

\begin{figure}[t]
	\begin{center}
		\includegraphics*{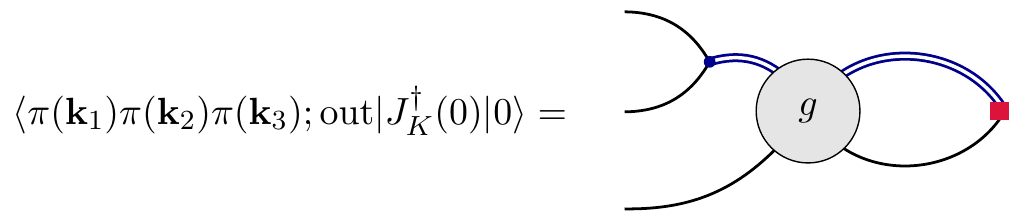}
	\end{center}
	\caption{Contributions to the infinite volume decay matrix element. The single and double lines denote the particle propagator and the full dimer propagator, respectively.}
	\label{fig:IV_decay_element}
\end{figure}

In the infinite volume the decay matrix element is given by, see Fig.~\ref{fig:IV_decay_element}:
\begin{align}\label{eq:IV_decay_element}
	\langle &\pi({\bf k}_1) \pi({\bf k}_2) \pi({\bf k}_3); \text{out} | J^\dagger_K(0) | 0 \rangle \nonumber\\[2mm]
	&=   4\pi \sum_{\ell m,\ell'm'} \frac{(-1)^\ell}{\sqrt{2\ell + 1}} \sum_{\alpha=1}^3 \int\frac{d^3k_{\perp}}{(2\pi)^3 2w_v(k)} \nonumber \\[2mm]
	 &\times f_{\ell'}((K-k_\alpha)^2) \mathscr{Y}_{\ell'm'}(\mathbf{\tilde k}_\alpha) g_{\ell'm', \ell m}(k_\alpha,k;K) \mathscr{Y}_{\ell, -m}(\underline{\bf k}) G_\ell((K-k)^2) \,.
\end{align}
Indeed this quantity is Lorentz invariant. The transformation property of $g_{\ell'm', \ell m}$ exactly cancels those of the spherical functions entering the expression.

\subsection{The three-particle LL formula}
At a given order $O(\epsilon^{2n})$ in the NREFT power counting, where $\epsilon$ is a generic small parameter and $p_\perp = O(\epsilon)$, only a finite number of effective decay couplings $G_i^{(\ell)}$ have to be taken into account. Noting that $\mathscr{Y}_{\ell, -m}(\underline{\bf k}) = O(\epsilon^\ell)$ and $\Delta_T = O(\epsilon^2)$, for a given angular momentum $\ell$ (even only):
\begin{equation}
  G_\ell(\Delta_T) = G_0^{(\ell)} + G_1^{(\ell)} \Delta_T + \dots
  + G_{n-\ell/2}^{(\ell)} \Delta_T^{n-\ell/2}
\end{equation}
Therefore, at order $O(\epsilon^{2n})$, there are $N = (n+1)(n+2)/2$ undetermined\footnote{Symmetry arguments could lower the number of independent couplings} couplings $G^{(\ell)}_i$.

Let $x_\alpha$ denote the finite volume decay matrix element extracted in some lattice setup $\alpha$, with total momentum $\mathbf{K} =  \mathbf{K}_\alpha$ and box length $L = L_\alpha$. Expanding the low-energy polynomial $G_\ell((K-k)^2)$, $x_\alpha$ can be written as a linear combination of the effective couplings:
\begin{equation}\label{eq:x_I}
  x_\alpha = L_\alpha^{3/2} \langle n | J^\dagger_K(0) | 0 \rangle_\alpha
  = \sum_{\ell=0}^{2n} \sum_{i=0}^{n-\ell/2}  a_{i}^{(\ell)}(K_\alpha, L_\alpha)
  G^{(\ell)}_i \equiv \sum_{I=1}^{N}   a_{\alpha I} \, G_I \,,
\end{equation}
where the amplitudes $a_{i}^{(\ell)}(K_\alpha, L_\alpha)$ can be read off from Eq.~\eqref{eq:FV_decay_element}. A similar expression can be found for the infinite volume decay matrix element:
\begin{equation}
  \langle \pi({\bf k}_1) \pi({\bf k}_2) \pi({\bf k}_3); \text{out} | J^\dagger_K(0)
  | 0 \rangle = \sum_{\ell=0}^{2n} \sum_{i=0}^{n-\ell/2}  A_{i}^{(\ell)}(K) G^{(\ell)}_i
  \equiv \sum_{I=1}^{N}  A_I(K) \, G_I  \,,
\end{equation}
where $A_{i}^{(\ell)}(K)$ can be read off from Eq.~\eqref{eq:IV_decay_element} and $K$ is the total momentum of the three-particle system. Measuring $N$ finite volume decay matrix elements in different lattice setups and using Eq.~\eqref{eq:x_I}, interpreted as a matrix equation, we can eliminate the dependence on the coupling constants $G_i^{\ell}$ of the infinite volume matrix element:
\begin{equation}\label{eq:LL_1}
	\langle \pi({\bf k}_1) \pi({\bf k}_2) \pi({\bf k}_3); \text{out} | J^\dagger_K(0) | 0 \rangle = \sum_{\alpha=1}^{N} (\Phi_3)_\alpha \cdot  L_\alpha^{3/2} |\langle n | J^\dagger_K(0) | 0 \rangle |_\alpha \,,
\end{equation}
where the LL-factor $(\Phi_3)_\alpha$ is given by
\begin{equation}\label{eq:LL_2}
	 (\Phi_3)_\alpha = \sum_{I=1}^{N} A_I(K) \, (a^{-1})_{I\alpha}
\end{equation}
and $a^{-1}$ is the inverse of the matrix
\begin{equation}\label{eq:LL_3}
	a = \begin{pmatrix}
	a^{(0)}_0(K_1, L_1) & a^{(0)}_1(K_1, L_1) & \dots &  a^{(0)}_N(K_1, L_1) & \dots & a^{(2n)}_0(K_1, L_1) \\
	\vdots && \ddots &&& \vdots \\
	a^{(0)}_0(K_N, L_N) & a^{(0)}_1(K_N, L_N) & \dots & a^{(0)}_N(K_N, L_N) & \dots & a^{(2n)}_0(K_N, L_N)
	\end{pmatrix}  \,,
\end{equation}
while 
\begin{equation}\label{eq:LL_4}
	A_I(K) = \begin{pmatrix}
		A^{(0)}_0(K) & A^{(0)}_1(K) & \dots & A^{(0)}_N(K) & \dots & A^{(2n)}_0(K)
	\end{pmatrix}  \,.
\end{equation}
In the equations above it is assumed that the box sizes $L_\alpha$ were adjusted, such that $K^0 = \sqrt{M_K^2 + \mathbf{K}_\alpha^2}$, for each $\alpha$.

The Eqs.~\eqref{eq:LL_1}-\eqref{eq:LL_4} display our finial result, resembling the LL equation for the three-particle sector. As in the two-particle sector, the LL factors merely depend on the short-range pion interactions. Fixing the parameters of the two- and three-particle scattering sector by a fit to the corresponding energy-spectra by using the quantization conditions fully determines the amplitudes $A^{(\ell)}_i(K)$ and $a^{(\ell)}_i(K,L)$. One then extracts the finite volume decay matrix element in various different moving frames. Here one has to ensure that the energy of the three-particle eigenstate coincides with an energy level of the moving kaon. This is done by choosing the box length $L$ appropriately. One can finally use Eqs.~\eqref{eq:LL_1}-\eqref{eq:LL_4} to obtain the infinite volume decay amplitude. Note also that the above expressions are very similar
in form to the three-particle LL equation derived in the RFT formalism
(see, e.g., Eq.~(2.43) in Ref.~\cite{Hansen:2021ofl}). For example,
the eigenvector $v$ in that paper can be related to the particle-dimer wave
function $\psi_n^{(\ell m)}$ from our paper and so on. We did not attempt,
however, to carry out a detailed comparison with Ref.~\cite{Hansen:2021ofl}.

Note that Eq.~\eqref{eq:LL_1} is manifestly relativistic-invariant. The infinite volume decay amplitude is a linear combination of invariant amplitudes $A^{(\ell)}_i(K)$.

\section{Conclusions}
\label{sec:concl}

\begin{itemize}
\item[i)] In the present paper we have derived a manifestly relativistic-invariant counterpart of the LL formula in the three-particle sector, describing the decay of a spinless particle into three likewise spinless, identical particles. Our result represents a generalization of the formula derived in Ref.~\cite{Muller:2020wjo}, which is reproduced at the leading order and at a vanishing total three-momentum. The importance of having an explicitly invariant framework lies, first and foremost, in the possibility to use the data from different moving frames for performing a global fit. This possibility is extremely valuable, especially beyond the leading order.

\item[ii)] A major technical modification in this paper consisted in the inclusion of all
  partial waves in the pair interactions as well as in the particle-dimer interactions.
  This renders the proof of the Lorentz-invariance a highly non-trivial issue, due to the
  necessity to take into account the Thomas-Wigner rotation. A solution
  of this problem has been found in the present paper.

\item[iii)] Another technical complication was related to the necessity of the diagonalization
  of the quantization condition in various irreps of the octahedral group and various subgroups thereof. In this paper we explicitly write down the quantization condition, projected onto different irreps, in the presence of an arbitrary number of partial waves.

\item[iv)] The LL factors, relating the infinite volume decay matrix elements to its counterparts in various different moving frames, depend on the local two- and three-particle interactions that emerge due to the rescattering in the final state. The corresponding parameters in the three-particle sector can be fixed by a fit to the energy spectrum.
Furthermore, as the quantization condition, derived in the present paper, is valid in an arbitrary moving frame, the energy spectra of the same lattice setups can be used, in which the finite volume decay amplitudes were determined.
	
\item[v)]
  The framework can be further generalized in a number of ways.
  Namely, in order to describe decays of particles with an arbitrary spin, the LL formula can be straightforwardly modified by an appropriate choice of source operators for the decaying particle and the three-particle final states, replacing $J^\dagger_K$ and $\mathscr{O}$ respectively. Furthermore, one may consider the case of the non-identical particles in the final state and include fermions
  (note that three pions carrying isospin have been already considered in Ref.~\cite{Hansen:2021ofl}).
  It can be expected that all these actions are pretty straightforward and can be performed by using the same methods as in the present article.
\end{itemize}

\begin{acknowledgments}
  The authors would like to thank H.-W. Hammer, M. Mai and F. Romero-Lopez
  	for interesting discussions. 
	The work of F.M. and A.R. was  funded in part by
	the Deutsche Forschungsgemeinschaft
	(DFG, German Research Foundation) – Project-ID 196253076 – TRR 110 and by the Ministry of Culture and Science of North Rhine-Westphalia through the NRW-FAIR project.
	A.R., in addition, thanks Volkswagenstiftung 
	(grant no. 93562) and the Chinese Academy of Sciences (CAS) President's
	International Fellowship Initiative (PIFI) (grant no. 2021VMB0007) for the
	partial financial support.
	The work of J.-Y.P. and J.-J.W. was supported by the National Natural Science Foundation of China (NSFC) under Grants No. 12135011, 12175239, 12221005, and by the National Key R\&D Program of China under Contract No. 2020YFA0406400. 
	
\end{acknowledgments}

\appendix

\section{A dimer field with the spin $\ell$}
\label{app:c}

We define a dimer field with the spin $\ell$ and projection $m=-\ell,\cdots,\ell$
from the tensor field $T_{\mu_{1}\cdots\mu_{\ell}}$. There are
three constraints on $T_{\mu_{1}\cdots\mu_{\ell}}$:
\begin{itemize}
\item Permutation symmetry in all indices
\begin{align}\label{eq:A1}
  T_{\mu_1\cdots\mu_i\cdots\mu_j\cdots\mu_\ell} &
  =T_{\mu_1\cdots\mu_j\cdots\mu_i\cdots\mu_\ell}\,,\quad\quad
  \mbox{for all }i,j\, .
\end{align}
\item The tensor $T_{\mu_{1}\cdots\mu_{\ell}}$ is traceless in each pair of indices
\begin{align}\label{eq:A2}
  g^{\mu_i\mu_j}
  T_{\mu_1\cdots\mu_i\cdots\mu_j\cdots\mu_\ell}
  & =0\, ,\quad\quad
  \mbox{for all }i,j\, .
\end{align}
\item For all $i$, the field obeys the constraint~(\ref{eq:A3}).
\end{itemize}
By considering the Lorentz transformation $\underline{\Lambda}$ that
boosts $v^{\mu}$ to $v_{0}^{\mu}=(1,\mathbf{0})$, one can define the
``rest-frame'' tensor
$\underline{T}^{\mu_{1}\cdots\mu_{\ell}}$ as
\begin{align}
  \underline{T}^{\mu_1\cdots\mu_\ell} & =\underline{\Lambda}_{\nu_1}^{\mu_1}
  \cdots\underline{\Lambda}_{\nu_\ell}^{\mu_\ell}T^{\nu_1\cdots\nu_\ell}.
\end{align}
Due to the constraint~(\ref{eq:A3}), the field
$\underline{T}^{\mu_1\cdots\mu_\ell}$
vanishes, when any of the indices is equal to $0$. For simplicity of the notations, we will use
spatial indices, $i_1,\cdots,i_\ell$, instead of the Lorentz
indices $\mu_1,\cdots,\mu_\ell$.

We begin with the case $\ell=1$, and write down
a transformation that relates the matrix $\underline{T}$ in the Cartesian
and spherical bases
\begin{align}
\underline{T}^{i} & =c_{1m}^{i}T_{1m}\, ,\quad\quad i=1,2,3,\quad m=-1,0,+1\, .
\end{align}
Here, $c_{1m}^{i}$ is given by
\begin{align}
c_{1,\pm1}^i=\frac{1}{\sqrt{2}}\begin{pmatrix}\mp1\\ 
-i\\
0
\end{pmatrix}_{\!i}, & \quad c_{1,0}^i=\begin{pmatrix}0\\
0\\
1
\end{pmatrix}_{\!i}.
\end{align}
The above result is obtained by postulating
\eq
\langle 0|\underline{T}^{i}|j\rangle=\delta_{ij}\, ,\quad\quad
\langle 0|T_{1m}|1n\rangle=\delta_{mn}\, ,
\en
and using the known relation between the basis vectors $|j\rangle$ and $|1n\rangle$ (the Condon-Shortley phase convention is adopted throughout this paper).

Inversely, one can find that
\begin{align}
T_{1m} & =\left(c^{-1}\right)_{1m}^{i}\underline{T}_{\,i},
\end{align}
where 
\begin{align}
\left(c^{-1}\right)_{1m}^{i} & =\left(c_{1m}^{i}\right)^{*}.
\end{align}
The spin-$\ell$ field is build up by the direct product of $\ell$
spin-1 fields, i.e., 
\begin{align}
T_{\ell m}=\mathscr{C}_{m_{1}\cdots m_{\ell}}^{\ell m}T_{1m_{1}}\otimes\cdots\otimes T_{1m_{\ell}} & =\mathscr{C}_{m_{1}\cdots m_{\ell}}^{\ell m}(c_{1m_{1}}^{i_{1}})^{*}\cdots(c_{1m_{\ell}}^{i_{\ell}})^{*}\underline{T}_{\,i_{1}}\otimes\cdots\otimes\underline{T}_{\,i_{\ell}}\nonumber \\[2mm]
 & \equiv\left(c^{-1}\right)_{\ell m}^{i_{1}\cdots i_{\ell}}\underline{T}_{\,i_{1}\cdots i_{\ell}}.
\end{align}
Here we are using the following notations:
\begin{align}
\left(c^{-1}\right)_{\ell m}^{i_{1}\cdots i_{\ell}} & =\mathscr{C}_{m_{1}\cdots m_{\ell}}^{\ell m}(c_{1m_{1}}^{i_{1}})^{*}\cdots(c_{1m_{\ell}}^{i_{\ell}})^{*},\\[2mm]
\underline{T}_{\,i_{1}\cdots i_{\ell}} & =\underline{T}_{\,i_{1}}\otimes\cdots\otimes\underline{T}_{\,i_{\ell}}.
\end{align}
The coefficient $\mathscr{C}_{m_{1}\cdots m_{\ell}}^{\ell m}$
for $m_1=\cdots=m_\ell=1$ 
can be read off directly, since the highest-weight state takes the form 
\begin{align}
  |\ell,\ell\rangle & =\underset{\ell}{\underbrace{|1,m_{1}=1\rangle\otimes\cdots\otimes|1,m_{\ell}=1\rangle}}\equiv
  \underset{\ell}{|\underbrace{1,\cdots,1}\rangle}.
\end{align}
From the above equation it follows that $\mathscr{C}_{1\cdots 1}^{\ell \ell}=1$.
For the lower-weight states, we can use the ladder operator
\begin{align}
J^{-}|j,m+1\rangle & =A_{j,m}^{-}|j,m\rangle,\quad A_{j,m}^{-}=\sqrt{(j-m)(j+m+1)}.
\end{align}
Acting on the state with maximum weight, one gets
\begin{align}
J^{-}|\ell,\ell\rangle & =A_{\ell,\ell-1}^{-}|\ell,\ell-1\rangle.
\end{align}
At the same time, we have, 
\begin{align}
J^{-}\underset{\ell}{|\underbrace{1,1,\cdots,1}\rangle} & =A_{1,0}\underset{\ell}{|\underbrace{0,1,\cdots,1}\rangle}+A_{1,0}\underset{\ell}{|\underbrace{1,0,\cdots,1}\rangle}+\cdots A_{1,0}\underset{\ell}{|\underbrace{1,1,\cdots,0}\rangle}\ .
\end{align}
This means that
\begin{align}
  \mathscr{C}_{0,1,\cdots,1}^{\ell,\ell-1} &
  =\mathscr{C}_{1,0,\cdots,1}^{\ell,\ell-1}=\cdots=\mathscr{C}_{1,1,\cdots,0}^{\ell,\ell-1}
  =\frac{A_{1,0}^{-}}{A_{\ell,\ell-1}^{-}}\, .
\end{align}
Continuing to act with the ladder operator, one gets
\begin{align}
\left(J^{-}\right)^{2}|\ell,\ell\rangle & =A_{\ell,\ell-1}A_{\ell,\ell-2}|\ell,\ell-2\rangle.
\end{align}
On the other hand, 
\begin{align}
 & \left(J^{-}\right)^{2}|1,1,\cdots,1\rangle\nonumber\\[2mm]
=\; & J^{-}\left(A_{1,0}^{-}|0,1,\cdots,1\rangle+A_{1,0}^{-}|1,0,\cdots,1\rangle+\cdots+A_{1,0}^{-}|1,1,\cdots,0\rangle\right)\nonumber\\[2mm]
=\; & \frac{2!}{\underbrace{1!1!0!\cdots 0!}_\ell}\,
\left(A_{1,0}^{-}\right)^{2}
\left(|0,0,\cdots,1,1\rangle+\cdots+|1,1,\cdots,0,0\rangle\right)\nonumber\\[2mm]
+&\; \frac{2!}{\underbrace{2!0!\cdots 0!}_\ell}\, A_{1,0}^{-}A_{1,-1}^{-}\left(|1,1,\cdots,-1\rangle+\cdots+|1,1,\cdots,-1\rangle\right)\, .
\end{align}
The first bracket in the above equation contains $\ell$ terms, in which
$-1$ stands in the first position, in the second position, and so on. The second
bracket contains $\frac{1}{2}\,\ell(\ell-1)$ terms, in which two zeros
stand in arbitrary positions. Hence,
\begin{align}
  |\ell,\ell-2\rangle  =&\frac{1}{A_{\ell,\ell-1}A_{\ell,\ell-2}}\Bigg(
  \frac{2!}{\underbrace{1!1!0!\cdots 0!}_\ell}\,
  \left(A_{1,0}^{-}\right)^{2}\Big[|0,0,1,\cdots,1,1\rangle
    +\cdots+|1,1,1,\cdots,0,0\rangle\Big]
  \nonumber \\[2mm]
  +&
  \frac{2!}{\underbrace{2!0!\cdots 0!}_\ell}\,\left(A_{1,0}^{-}A_{1,-1}^{-}\right)\Big[|-1,1,\cdots,1\rangle
    +\cdots+|1,1,\cdots,-1\rangle\Big]\Bigg)
    \, .
\end{align}
From this, we can read off that
\begin{align}
\mathscr{C}_{0,0,1,\cdots,1,1}^{\ell,\ell-2} 
=\mathscr{C}_{1,0,0,\cdots,1,1}^{\ell,\ell-2}=\cdots=\mathscr{C}_{1,1,\cdots,0,0}^{\ell,\ell-2}
&=\frac{2!}{\underbrace{1!1!0!\cdots 0!}_\ell}\,
\frac{\left(A_{1,0}^{-}\right)^{2}}{A_{\ell,\ell-1}A_{\ell,\ell-2}}\, ,
\nonumber\\[2mm]  
  \mathscr{C}_{-1,1,\cdots,1}^{\ell,\ell-2} 
  =\mathscr{C}_{1,-1,\cdots,1}^{\ell,\ell-2}=\cdots=\mathscr{C}_{1,1,\cdots,-1}^{\ell,\ell-2}&=\frac{2!}{\underbrace{2!0!\cdots 0!}_\ell}\,
  \frac{A_{1,0}^{-}A_{1,-1}^{-}}{A_{\ell,\ell-1}A_{\ell,\ell-2}}\, .
\end{align}
This procedure can be continued in a straightforward manner. Recalling that
$A_{1,-1}^{-}=A_{1,0}^{-}=\sqrt{2}$, generic expression for the coefficients
$\mathscr{C}$ is given by
\begin{align}
  \mathscr{C}_{m_{1}\cdots m_{\ell}}^{\ell m} &
  =\frac{(\sqrt{2})^{\ell-m}}{A_{\ell,\ell-1}^{-}\cdots A_{\ell,m}^{-}}\,
\frac{(\ell-m)!}{(1-m_1)!\cdots (1-m_\ell)!}\, .
\end{align}
To summarize, $T_{\ell m}$ and $\underline{T}^{\mu_{1}\cdots\mu_{\ell}}$ are related by
\begin{align}
T_{\ell m} & =\left(c^{-1}\right)_{\ell m}^{\mu_{1}\cdots\mu_{\ell}}\underline{T}_{\,\mu_{1}\cdots\mu_{\ell}}\, ,
\end{align}
and, inversely, 
\begin{align}
\underline{T}^{\mu_{1}\cdots\mu_{\ell}} & =c_{\ell m}^{\mu_{1}\cdots\mu_{\ell}}T_{\ell m},\quad\text{(}\ell\text{ not summed)}
\end{align}
Here the matrix $c$ matrix is given by 
\begin{align}\label{eq:cases}
c_{\ell m}^{\mu_{1}\cdots\mu_{\ell}} & =\begin{cases}
\mathscr{C}_{m_{1}\cdots m_{\ell}}^{\ell m}(c_{1m_{1}}^{i_{1}})\cdots(c_{1m_{\ell}}^{i_{\ell}}), & (\mu_{1}=i_{1},\cdots,\mu_{\ell}=i_{\ell})\, ,\\
0, & \mbox{if at least one of $\mu_i=0$}\, .
\end{cases}
\end{align}
The matrix $c^{-1}$ is complex conjugate of $c$:
\begin{align}
\left(c^{-1}\right)_{\ell m}^{\mu_{1}\cdots\mu_{\ell}} & =\left(c_{\ell m}^{\mu_{1}\cdots\mu_{\ell}}\right)^{*}\, .
\end{align}
Moreover, it obeys the constraints that are imposed by Eqs.~(\ref{eq:A1})-(\ref{eq:A2}), namely
\begin{align}
c_{\ell m}^{\mu_{1}\cdots\mu_{i}\cdots\mu_{j}\cdots\mu_{\ell}} & =c_{\ell m}^{\mu_{1}\cdots\mu_{j}\cdots\mu_{i}\cdots\mu_{\ell}},\\[2mm]
g_{\mu_i\mu_j}c_{\ell m}^{\mu_{1}\cdots\mu_i\cdots\mu_j\cdots\mu_{\ell}} & =0.
\end{align}
The constraint from Eq.~(\ref{eq:A3}) is obeyed automatically,
owing to Eq.~(\ref{eq:cases}).

\section{The dimer propagator in a finite volume}
\label{app:dimer_prop}

In this appendix, we briefly sketch the calculation of the
finite-volume self energy, displayed in Eq.~(\ref{eq:Sigma}). The on-shell
momenta in Eq.~(\ref{eq:momenta}) are defined as
\eq
\hat q_\mu=q_\mu-v_\mu(vq-w_v(q))\, ,\quad\quad
     {\hat q'}_\mu=(P-q)_\mu-v_\mu(v(P-q)-w_v(P-q))\, .
     \en
     Furthermore, for any four-momentum,
\eq
     \tilde p=\Lambda(v_0,v)\Lambda(v,u)\hat p
     =\left(\Lambda(v_0,v)\Lambda(v,u)\Lambda^{-1}(v_0,u)\right)\Lambda(v_o,u)\hat p\, .
     \en
The product of three matrices in the brackets is a pure rotation:
\eq
\left(\Lambda(v_0,v)\Lambda(v,u)\Lambda^{-1}(v_0,u)\right)^{\mu\nu}=R^{\mu\nu}(v,u)\, ,
\en
where
\eq
R^{00}&=&1\, ,\quad\quad R^{0i}=R^{i0}=0\, ,
\nonumber\\[2mm]
R^{ij}&=&-\delta^{ij}-\frac{(1-v^0)u^iu^j}{(1+(uv))(1+u^0)}
-\frac{(1-u^0)v^iv^j}{(1+(uv))(1+v^0)}
-\frac{{\bf u}{\bf v}(v^iu^j+u^iv^j)}{(1+(uv))(1+u^0)(1+v^0)}
\nonumber\\[2mm]
&+&\frac{(1+u^0+v^0+uv)(v^iu^j-u^iv^j)}{(1+(uv))(1+u^0)(1+v^0)}\, .
\en
Below the elastic threshold $P^2<4m^2$, one can merely replace the sum by an integral
in Eq.~(\ref{eq:Sigma2}) -- the corrections to the infinite-volume limit are exponentially suppressed. Therefore, we concentrate on the case $P^2\geq 4m^2$ here.
In this case, the expression for the self-energy can be further simplified.
Namely, the quantities $\hat q,\hat q'$ in the numerator in Eq.~(\ref{eq:Sigma}) can be expanded in Taylor series
in $(vq-w_v(q))$ and $(v(P-q)-w_v(P-q))$.
All terms, except the first one, obviously vanish in this expansion, since after the
integration over $q^0$ one gets
a low-energy polynomial and the sum vanishes, if the dimensional regularization
is used (we remind the reader that, by implicit convention, the infinite-volume limit
is already subtracted in this sum).
This expansion, in particular,
leads to the replacement $u^\mu\to P^\mu/\sqrt{s}$, where
$P^\mu=(\sqrt{s+{\bf P}^2},{\bf P})$. Hence, $u^\mu$ becomes independent of
the summation momentum and, after integration over $q^0$,
the expression of the self-energy simplifies to
\eq
\Sigma^L_{\ell' m',\ell m}(P)=\sum_{m'''m''}\mathscr{D}^{(\ell')}_{m'm'''}(R(u,v))
\bar\Sigma^L_{\ell' m''',\ell m''}(P)\left(\mathscr{D}^{(\ell')}_{mm''}(R(u,v))\right)^*\, .
\en
where
 \eq\label{eq:Sigma2}
\bar\Sigma^L_{\ell' m',\ell m }(P)=f_{\ell'}(P^2)f_\ell(P^2)
\frac{1}{2L^3}\sum_{\bf q}\frac{\left(\mathscr{Y}_{\ell' m'}({\bf k})\right)^*\mathscr{Y}_{\ell m}({\bf k})}
     {2w({\bf q})2w({\bf P}-{\bf q})(w({\bf q})+w({\bf P}-{\bf q})-P^0)}\, ,\quad\quad
     \en
     where
     \eq
        {\bf k}={\bf l}-{\bf u}\frac{{\bf l}{\bf u}}{u^0(u^0+1)}\, ,\quad\quad
        {\bf l}={\bf q}-\frac{1}{2}\,{\bf P}\, ,
        \en
At this point it is seen that the dependence of $v^\mu$ is trivially factored out and is contained in the Wigner $D$-functions only.

        Next, we shall use the well-known addition property of the spherical functions
           \eq
           \mathscr{Y}^*_{\ell' m'}({\bf k})\mathscr{Y}_{\ell m }({\bf k})
           &=&(-1)^{m'}\sqrt{\frac{(2\ell'+1)(2\ell+1)}{4\pi}}
\nonumber\\[2mm]
           &\times&\sum_{js}(-1)^j\sqrt{2j+1}
|{\bf k}|^{\ell'+\ell-j}           
           \begin{pmatrix}
             \ell' & \ell & j\cr
             -m' & m & -s
             \end{pmatrix}
                \begin{pmatrix}
             \ell' & \ell & j\cr
             0 & 0 & 0
             \end{pmatrix}
                \mathscr{Y}_{js}({\bf k})\, .
                \en
Here,  $\begin{pmatrix} \ell' & \ell & j\cr -m' & m & -s \end{pmatrix}$ and
$\begin{pmatrix} \ell' & \ell & j\cr0 & 0 & 0 \end{pmatrix}$
denote the Wigner $3j$ symbols.
Note that $\ell'+\ell-j$ should be an even integer -- otherwise,
the second of the $3j$ symbols would vanish.

Furthermore, we shall use the following identity (see, e.g.,~\cite{Bernard:2012bi}):
\eq
\frac{1}{2w({\bf q})2w({\bf P}-{\bf q})(w({\bf q})+w({\bf P}-{\bf q})-P^0)}
=\frac{1}{2P^0}\,\frac{1}{{\bf l}^2-\dfrac{({\bf l}{\bf P})^2}{{P^0}^2}-q_0^2}+\cdots\, ,
\en
where $q_0^2=\sqrt{\dfrac{s}{4}-m^2}$ and the ellipses stand for the terms
that are low-energy polynomials and thus do not contribute.

It is immediately seen that the denominator in the r.h.s
of the above relation is equal to ${\bf k}^2-q_0^2$. Taking into account
that  $\ell'+\ell-j$ is even, Eq.~(\ref{eq:Sigma2}) can be rewritten as follows
 \eq\label{eq:Sigma3}
 \bar\Sigma^L_{\ell' m',\ell m}(P)&=&\frac{\pi^2}{LP^0}\,f_{\ell'}(P^2)f_\ell(P^2)q_0^{\ell'+\ell-j} 
 \sqrt{\frac{(2\ell'+1)(2\ell+1)}{4\pi}}
 \nonumber\\[2mm]
&\times&\sum_{js}(-1)^{j+m}\sqrt{2j+1}           
           \begin{pmatrix}
             \ell' & \ell & j\cr
             -m' & m & -s
             \end{pmatrix}
                \begin{pmatrix}
             \ell' & \ell & j\cr
             0 & 0 & 0
             \end{pmatrix} Z^{\bf d}_{js}(1;s)\, ,
                \en
where
\eq
Z^{\bf d}_{js}(1;s)
&=&\sum_{{\bf r}\in P_d}\frac{\mathscr{Y}_{js}({\bf r})}{{\bf r}^2-\eta^2}\, \quad\quad\, ,
\nonumber\\[2mm]
P_d&=&\left\{
{\bf r}\in \mathbb{R}^3|r_\parallel=\gamma^{-1}(n_\parallel-\frac{1}{2}\,|{\bf d}|),
~{\bf r}_\perp={\bf n}_\perp,~{\bf n}\in\mathbb{Z}^3\right\}\, ,
\en
and
\eq
   {\bf d}=\frac{L}{2\pi}\,{\bf P}\, ,\quad\quad
   \eta=\frac{L}{2\pi}\,q_0\, ,\quad\quad
   \gamma=\frac{P^0}{\sqrt{s}}\, .
   \en
   Final remark is in order. As mentioned already, our calculations concern the finite-volume corrections only. The infinite-volume part has to be added by hand
   at the end of the day. In our case, the real part of the loop function
   $J(s)$ should be added, see Eq.~(\ref{eq:IJ}). Once this is done,
   the finite-volume loop function below threshold smoothly converges
   to the infinite-volume result, when $L\to\infty$.
   
\section{Lorentz transformations for $Z_{\sf loc}$}
\label{app:Zloc}

In order to establish the properties of $Z_{\sf loc}$ under Lorentz transformations,
we shall use the following well-known properties of the Wigner functions:
\eq\label{eq:properties}
\mathscr{D}^{(j)}_{mn}(R)\mathscr{D}^{(j')}_{m'n'}(R)
&=&\sum_{J=|j-j'|}^{j+j'}
\langle jm,j'm'|J(m+m')\rangle
\langle jn,j'n'|J(n+n')\rangle
\mathscr{D}^{(J)}_{(m+m')(n+n')}(R)\, ,
\nonumber\\[2mm]
\mathscr{D}^{(j)}_{mm'}(R)&=&(-1)^{m-m'}\left(
  \mathscr{D}^{(j)}_{(-m)(-m')}(R)\right)^*\, . 
\en
Now, taking into account the fact that $T_{JL'L}$ are invariant under Lorentz transformations, one may write
\eq
\left(Z_{\sf loc}\right)_{\ell'm',\ell m}(\Omega p,\Omega q)&=&
4\pi\sum_{JLL'}\sum_{MM'}T_{JLL'}\delta_{M'M}
\langle L'(M'-m'),\ell'm'|JM'\rangle\mathscr{Y}_{L'(M'-m')}(R\underline{\bf p})
\nonumber\\[2mm]
&\times&\langle L(M-m),\ell m|JM\rangle\left(\mathscr{Y}_{L(M-m)}(R\underline{\bf q})\right)^*\, .
\en
In the above equation, one may use the transformation law, given in
Eq.~(\ref{eq:Wignerd}). In addition, one may replace the Kronecker-delta
$\delta_{M'M}$ by
\eq
\delta_{M'M}=\sum_N\mathscr{D}^{(J)}_{M'N}(R)\left(\mathscr{D}^{(J)}_{MN}(R)\right)^*\, .
\en
Using now the second equation in (\ref{eq:properties}), the local contribution can be rewritten as follows
\eq
&&\left(Z_{\sf loc}\right)_{\ell'm',\ell m}(\Omega p,\Omega q)\,=\,
4\pi\sum_{JLL'}\sum_{MM'}T_{JLL'}\sum_N\sum_{nn'}\mathscr{D}^{(J)}_{M'N}(R)\left(\mathscr{D}^{(J)}_{MN}(R)\right)^*
\nonumber\\[2mm]
&&\quad\quad\quad\times\,\langle L'(M'-m'),\ell'm'|JM'\rangle\mathscr{Y}_{L'n'}(\underline{\bf p})
(-1)^{M'-m'-n'}\mathscr{D}^{(L')}_{(m'-M')(-n')}(R)
\nonumber\\[2mm]
&&\quad\quad\quad\times\,\langle L(M-m),\ell m|JM\rangle\left(\mathscr{Y}_{Ln}(\underline{\bf q})\right)^*
(-1)^{M-m-n}\left(\mathscr{D}^{(L)}_{(m-M)(-n)}(R)\right)^*
\, .
\en
Next, one uses addition theorem, given in Eq.~(\ref{eq:properties}) and performs the sum
over $M',M$ afterwards. The addition theorem gives
\eq
&&\mathscr{D}^{(L')}_{(m'-M')(-n')}(R)\mathscr{D}^{(J)}_{M'N}(R)
\nonumber\\[2mm]
&&\quad\quad =\,\sum_{j'}
\langle L'(m'-M'),JM'|j'm'\rangle
\langle L'(-n'),JN|j'(N-n')\rangle
\mathscr{D}^{(j')}_{m'(N-n')}(R)\, ,
\en
and the same for the ``unprimed'' indices. Furthermore, the summation with $M',M$
is carried out by
using the symmetry properties for the Clebsch-Gordan coefficients. For example,
\eq
&&\sum_{M'}(-1)^{M'-m'-n'}\langle L'(m'-M'),JM'|j'm'\rangle\langle L'(M'-m'),\ell m'|JM'\rangle
\nonumber\\[2mm]
&=&\sum_{M'}(-1)^{J-n'-\ell'}\sqrt{\frac{2J+1}{2\ell'+1}}\langle L'(m'-M'),JM'|j'm'\rangle\langle L'(m'-M'),JM'|\ell m'\rangle
\nonumber\\[2mm]
&=&(-1)^{J-n'-\ell'}\sqrt{\frac{2J+1}{2\ell'+1}}\delta_{j'\ell'}\, ,
\en
and, similarly, for the ``unprimed'' indices.
Substituting this expression into the original
formula gives
\eq
&&\left(Z_{\sf loc}\right)_{\ell'm',\ell m}(\Omega p,\Omega q)\,=\,
4\pi\sum_{JLL'}\sum_{n'n}\sum_{j'j}\sum_N T_{JLL'}
\nonumber\\[2mm]
&&\quad\times\,(-1)^{-n'+J-\ell'}\sqrt{\frac{2J+1}{2\ell'+1}}\delta_{j'\ell'}
\langle L'(-n'),JN|j'(N-n')\rangle \mathscr{Y}_{L'n'}(\underline{\bf p})
\mathscr{D}^{(j')}_{m'(N-n')}(R)
\nonumber\\[2mm]
&&\quad\times\,(-1)^{-n+J-\ell}\sqrt{\frac{2J+1}{2\ell+1}}\delta_{j\ell}
\langle L(-n),JN|j(N-n)\rangle \left(\mathscr{Y}_{Ln}(\underline{\bf q})\right)^*
\left(\mathscr{D}^{(j)}_{m(N-n)}(R)\right)^*\, .\quad\quad
\en
Using the symmetries of the Clebsch-Gordan coefficients again, we get:
\eq
\left(Z_{\sf loc}\right)_{\ell'm',\ell m}(\Omega p,\Omega q)&=&
4\pi\sum_{JLL'}\sum_{n'n}\sum_N T_{JLL'}
\nonumber\\[2mm]
&\times&
\langle L'n',\ell'(N-n')|JN\rangle \mathscr{Y}_{L'n'}(\underline{\bf p})
\mathscr{D}^{(j')}_{m'(N-n')}(R)
\nonumber\\[2mm]
&\times&
\langle Ln,\ell(N-n)|JN\rangle \left(\mathscr{Y}_{Ln}(\underline{\bf q})\right)^*
\left(\mathscr{D}^{(j)}_{m(N-n)}(R)\right)^*\, .
\en
Finally, one arrives at
\eq
\left(Z_{\sf loc}\right)_{\ell'm',\ell m}(\Omega p,\Omega q)=
\sum_{m'''m''}\mathscr{D}^{(\ell')}_{m'm'''}(R)
\left(Z_{\sf loc}\right)_{\ell'm''',\ell m''}(p,q)
\left(\mathscr{D}^{(\ell)}_{mm''}(R)\right)^*\, .
\en

   \section{Projection onto the various irreps}
   \label{app:cubic}

   In this appendix, we give a detailed derivation of Eqs.~(\ref{eq:diagonal}) and
   (\ref{eq:AGamma}). To this end, we shall transform Eq.~(\ref{eq:projection}) by using the transformation property of $\mathscr{A}$, manifested in Eq.~(\ref{eq:transformation}):
   \eq\label{eq:projection1}
&&\mathscr{A}_{\sigma'(t'\Delta')\Sigma',\sigma(t\Delta)\Sigma}^{\ell'\Gamma'\alpha',\ell\Gamma\alpha}(p_r,q_s)\,
=\,\frac{s_{\sigma'}}{G}\,\frac{s_\Sigma}{G}\,
\sum_{g',g\in{\cal G}}\sum_{\lambda'\rho',\lambda\rho}\sum_{\lambda'''\lambda''}
\langle\Sigma'\rho',\Delta'\lambda'|\Gamma'\alpha'\rangle T^{(\Sigma')}_{\rho'\sigma'}(g')
T^{(\Delta')}_{\lambda'\lambda'''}(g)
\nonumber\\[2mm]
&&\quad\quad\times\, 
\mathscr{A}_{\lambda'''(t'\Delta'),\lambda''(t\Delta)}^{\ell'\ell}(\underbrace{g^{-1}g'}_{=g''}p_r,q_s)T^{(\Delta)}_{\lambda''\lambda}(g^{-1})
T^{(\Sigma)}_{\sigma\rho}(g^{-1})
\langle\Sigma\rho,\Delta\lambda|\Gamma\alpha\rangle
\nonumber\\[2mm]
&&
\quad\quad=\, \frac{s_{\Sigma'}}{G}\,\frac{s_\Sigma}{G}\,
\sum_{g, g''\in{\cal G}}\sum_{\lambda'\rho',\lambda\rho}
\sum_{\lambda'''\lambda''\sigma''}
\langle\Sigma'\rho',\Delta'\lambda'|\Gamma'\alpha'\rangle
T^{(\Sigma')}_{\rho'\sigma''}(g)T^{(\Sigma')}_{\sigma''\sigma'}(g'')
T^{(\Delta')}_{\lambda'\lambda'''}(g)
\nonumber\\[2mm]
&&\quad\quad\times\, 
\mathscr{A}_{\lambda'''(t'\Delta'),\lambda''(t\Delta)}^{\ell'\ell}
(g''p_r,q_s)T^{(\Delta)}_{\lambda''\lambda}(g^{-1})
T^{(\Sigma)}_{\sigma\rho}(g^{-1})
\langle\Sigma\rho,\Delta\lambda|\Gamma\alpha\rangle\, .
\en
In what follows, we shall use the relation
\eq
&&\sum_{\rho'\lambda'}\langle\Sigma'\rho',\Delta'\lambda'|\Gamma'\alpha'\rangle
T^{(\Sigma')}_{\rho'\sigma''}(g)
T^{(\Delta')}_{\lambda'\lambda'''}(g)
\nonumber\\[2mm]
&&\quad\quad=\sum_{\rho'\lambda'}\langle\Sigma'\rho',\Delta'\lambda'|\Gamma'\alpha'\rangle
\sum_{\Xi\beta'\gamma'}\langle\Sigma'\rho',\Delta'\lambda'|\Xi'\beta'\rangle
\langle\Sigma'\sigma'',\Delta'\lambda'''|\Xi'\gamma'\rangle T^{(\Xi')}_{\beta'\gamma'}(g)
\nonumber\\[2mm]
&&\quad\quad=\sum_{\gamma'}\langle\Sigma'\sigma'',\Delta'\lambda'''|\Gamma'\gamma'\rangle T^{(\Gamma')}_{\alpha'\gamma'}(g)\, .
\en
A similar relation holds also for the product $T^{(\Sigma)}(g^{-1})\times T^{(\Delta)}(g^{-1})$. With the use of these relations, the original expression simplifies to
\eq
&&\mathscr{A}_{\sigma'(t'\Delta')\Sigma',\sigma(t\Delta)\Sigma}^{\ell'\Gamma'\alpha',\ell\Gamma\alpha}(p_r,q_s)\,
=\,\frac{s_{\sigma'}}{G}\,\frac{s_\Sigma}{G}\,
\sum_{g,g''\in{\cal G}}\sum_{\gamma\gamma'}\sum_{\lambda'''\lambda''\sigma''}
\langle\Sigma'\sigma'',\Delta'\lambda'''|\Gamma'\gamma'\rangle
T^{(\Gamma')}_{\alpha'\gamma'}(g)
\nonumber\\[2mm]
&&\quad\quad\times\, T^{(\Sigma')}_{\sigma''\sigma'}(g'')
\mathscr{A}_{\lambda'''(t'\Delta'),\lambda''(t\Delta)}^{\ell'\ell}(g''p_r,q_s)
\langle \Sigma\sigma,\Delta\lambda''|\Gamma\gamma\rangle
T^{(\Gamma)}_{\gamma\alpha}(g^{-1})\, .
\en
Carrying out the summation over $g$ with the use of the orthogonality condition of the matrices of the irreps, we finally arrive at Eqs.~(\ref{eq:diagonal}) and (\ref{eq:AGamma}).

\bibliographystyle{unsrt}
\bibliography{references}

\end{document}